\documentclass[prd,aps,showpacs,tightenlines,nofootinbib,preprint,amssymb]{revtex4}
\usepackage{color}
\input{colordvi.tex}
\usepackage{amsmath}
\usepackage[dvips]{graphicx}
\usepackage{subfigure}
\newcommand{\beq}{\begin{equation}}
\newcommand{\eeq}{\end{equation}}
\newcommand{\beqa}{\begin{eqnarray}}
\newcommand{\eeqa}{\end{eqnarray}}
\newcommand{\simg}{\gtrsim}

\newcommand{\CO}{{\cal O}}
\newcommand{\vecs}[1]{\mbox{\boldmath${#1}$}}

\def\be{\bar{e}}
\def\bu{\bar{u}}
\def\bd{\bar{d}}

\begin{document}
\widetext
\draft

\title{
Gravitational Waves from $Q$-ball Formation}

\author{Takeshi Chiba}%
\affiliation{Department of Physics, College of Humanities and Sciences,
Nihon University, Tokyo 156-8550, Japan}
\author{Kohei Kamada}%
\affiliation{Department of Physics, Graduate School of Science,
The University of Tokyo, Tokyo 113-0033, Japan}
\affiliation{Research Center for the Early Universe (RESCEU),
Graduate School of Science, The University of Tokyo, Tokyo 113-0033, Japan}
\author{Masahide Yamaguchi}%
\affiliation{Department of Physics and Mathematics, Aoyama Gakuin
University, Sagamihara 229-8558, Japan}

\date{\today}

\pacs{98.80.Cq ; 04.30.Db }

\begin{abstract}
We study the detectability of the gravitational waves (GWs) from the
$Q$-ball formation associated with the Affleck-Dine (AD) mechanism, taking
into account both the dilution effects due to $Q$-ball domination and to
finite temperature.  The AD mechanism predicts the formation of
nontopological solitons, $Q$-balls, from which GWs are generated.
$Q$-balls with large conserved charge $Q$ can produce a large amount of GWs. 
On the other hand, the decay rate of such $Q$-balls is so small that they
may dominate the energy density of the Universe, which implies that GWs
are significantly diluted and that their frequencies are redshifted
during the $Q$-ball dominated era. Thus, the detectability of the GWs
associated with the formation of $Q$-balls is determined by these two
competing effects. We find that there is a finite but small parameter 
region where such GWs may be detected by future detectors such as DECIGO 
or BBO, only in the case when the thermal logarithmic potential dominates 
the potential of the AD field. Otherwise GWs from $Q$-balls would not be 
detectable even by these futuristic detectors: $\Omega_{\rm GW}^0<10^{-21}$.
Unfortunately, for such parameter region the present baryon asymmetry 
of the Universe can hardly be explained unless one fine-tunes $A$-terms 
in the potential. However the detection of such a GW background may give 
us an information about the early Universe, for example, it may suggest 
that the flat directions with $B-L=0$ are favored.
\end{abstract}

\maketitle
\vspace{1.5cm}

\section{Introduction}

Primordial gravitational waves (GWs) provide us with a lot of important
information about the early Universe because the interaction of GWs with matter is very weak and 
they carry the memory of cosmic history during and after inflation 
\cite{review}. One of the main sources of such primordial GWs is
cosmic inflation \cite{Starobinsky:1979ty}, during which stochastic GWs
are generated with a nearly scale-invariant spectrum with frequencies 
ranging from $10^{-15}$ Hz to $10^{5}$ Hz, depending on inflation
models. Though the amplitude is typically very small, next-generation
gravitational detectors like DECIGO \cite{DECIGO} and BBO \cite{BBO} as
well as indirect observations through the $B$-modes in the cosmic microwave
background (CMB) anisotropy \cite{cmbpol} might be able to detect such
GWs.  The detection of such stochastic GWs can reveal the history of the
early Universe such as the changes of the number of the massless degree
of freedom \cite{Watanabe:2006qe}, the reheating temperature
\cite{Nakayama:2008ip}, the neutrino decoupling \cite{Weinberg:2003ur},
the lepton asymmetry \cite{Ichiki:2006rn}, and so on. Additional GWs can
also be generated from cosmological mechanisms like (local/global) phase
transitions \cite{phasetransition} and preheating after inflation
\cite{preheating,garcia-bellido} or astrophysical origins like the collapse of
Population III stars \cite{Buonanno:2004tp}.

Recently, yet another interesting mechanism to produce GWs was proposed
in Ref. \cite{Kusenko:2008zm}, in which it is shown that significant
GWs are emitted during the $Q$-ball formation associated with the Affleck-Dine
(AD) mechanism of baryogenesis.
In this scenario, the angular momentum, the rotation around the origin
of scalar fields that carry the baryon or lepton number, is dynamically
generated, which implies that the baryon or lepton asymmetry is
produced. When initially almost homogeneous scalar fields start to
rotate, they suffer from spatial instabilities, and their fluctuations
begin to grow if the scalar field potential driving the field rotation
is flatter than the quadratic potential. Such fluctuations finally
settle down into nontopological solitons called $Q$-balls, whose
existence and stability are guaranteed by a conserved charge, $Q$,
associated with a global symmetry \cite{Coleman}.  Since the formation process
of such $Q$-balls is inhomogeneous and not spherical, GWs can be
generated during the formation.

In order to calculate the present amplitude of such GWs, one has to
estimate not only the amount of GWs at the formation of $Q$-balls but also
the dilution factor during the cosmic history after
the production of the GWs.  As shown in Ref. \cite{Kusenko:2008zm}, the
energy density of the GWs at the $Q$-ball formation is proportional to
some powers of the field value of the Affleck-Dine condensate, which
implies that the initial energy density of the GWs becomes large if the
typical charge $Q$ of $Q$-balls is large. On the other hand, the lifetime
of $Q$-balls becomes longer for larger $Q$ because the temperature at the
decay of $Q$-balls is typically proportional to the inverse square root of
the charge $Q$ \cite{Kawasaki:2006yb}. Therefore, $Q$-balls with large $Q$
can quickly dominate the energy density of the Universe and hence dilute
the GWs significantly. Thus, the maximum value of the present amplitude
of such GWs is determined by the balance of the above two competing
effects. In Refs.  \cite{Kusenko:2008zm}, in order to avoid the large
dilution by the $Q$-ball dominance, $Q$-balls are assumed to decay quickly
by some artificial effects, which are unnatural in the context of the
minimal supersymmetric standard model (MSSM).  Then, in this paper, we
reconsider the decay of $Q$-balls without such effects and calculate the
amplitude of GWs from $Q$-balls taking into account of the dilution factor
correctly,
which results in dramatic changes in the present amplitude of the GWs from the $Q$-ball formation.

The properties of the Affleck-Dine mechanism \cite{AD} and the 
subsequent formation of $Q$-balls 
depend on the supersymmetry (SUSY) breaking mechanism
\cite{Kusenko97,Enqvist98}. This is mainly because the effective
potential of a flat direction and the gravitino mass are quite different
for different mediation mechanisms. There are many types of $Q$-balls such
as gauge-mediation type \cite{Kusenko97,kk99}, gravity-mediation type
\cite{Enqvist98,kk00a}, new type \cite{kk00b}, delayed type \cite{kk01},
and so on.

Moreover, the effective potential of a flat direction consists of
thermal terms as well as the zero-temperature terms. The properties of
$Q$-balls are quite different depending on which term dominates the
potential energy. When thermal (logarithmic) effects \cite{Anisimov}
dominate the effective potential, the formed $Q$-balls, called thermal log
type \cite{kk01}, have an interesting property. Namely, while the energy
density of other types of $Q$-balls decreases like matter, that of the thermal
log type $Q$-ball decreases at least as rapid as radiation. This is
because the thermal logarithmic potential itself also decreases with the
cosmic expansion while the number of $Q$-balls in a comoving volume does
not change.  Hence, this type of $Q$-balls cannot dominate the energy
density of the Universe and do not dilute GWs. This is favorable for the
detection of the GWs from the $Q$-ball formation because the dilution
during the $Q$-ball dominated era is the main obstacle for the detection.
In fact, as is explicitly shown in Appendix \ref{App:zero}, GWs from
the $Q$-ball formation in the zero-temperature potential may not be
detectable even by the next-generation detectors. Therefore, in this
paper, we concentrate on the thermal log type $Q$-balls and estimate the
present amplitudes and frequencies of the GWs at the formation of such
$Q$-balls. We show that such GWs may be detected by the next-generation
gravitational detectors like DECIGO and BBO if particular conditions of
reheating temperature, the initial field value of the AD field,
gravitino mass and messenger mass are realized in the gauge-mediated
SUSY-breaking model.  However, we also find that such a condition spans
a very small region in the parameter space.  Moreover, we also find that
 it is difficult to
explain the present baryon asymmetry for such a parameter region,  
unless one fine-tunes the $CP$-violating $A$-terms in the potential.

The paper is organized as follows. In Sec. \ref{sec:AD}, we give a brief
review of the dynamics of the Affleck-Dine baryo/leptogenesis and the
properties of the subsequently produced $Q$-balls, particularly paying
attention to the decay of $Q$-balls. In Sec. \ref{sec:GWs}, we evaluate
how many GWs are produced at the formation of $Q$-balls and are diluted
during the cosmic history, which yield the present amplitude of the GWs
{}from the $Q$-ball formation. Section \ref{sec:DC} is devoted to discussion
and conclusions. We concentrate on thermal log type $Q$-balls in the main
body of the paper. The cases with other types of $Q$-balls, in which the
zero-temperature effect (and the negative thermal log effect)
dominate the effective potential are discussed in Appendix
\ref{App:zero}.

\section{Affleck-Dine mechanism and $Q$-balls}

\label{sec:AD}

In this section, we briefly review the AD mechanism, and the formation
and the fate of the nontopological solitons, $Q$-balls.  In
supersymmetric theories, there are many flat directions along which the
scalar potentials become flat in the global SUSY limit, and hence such
scalar fields can easily acquire large field values. The flat directions
in the MSSM consist of squarks, sleptons, and Higgs fields, and are
parameterized by composite gauge-invariant monomial operators such as
$\bu\bd\bd$ and $LL\be$ \cite{Gherghetta}. Thus, flat directions can
carry baryon ($B$) or lepton ($L$) charges in general. This is
the reason why the AD mechanism can be one of the powerful models of
baryogenesis. We will see its details in the following subsection.

The dynamics of a flat direction can be expressed in terms of a scalar
field $\Phi$ (the AD field).  We consider the dynamics of a flat
direction $\Phi$, taking into account the finite temperature effects
since these effects play important roles in the formation and the
evolution of $Q$-balls.

\subsection{Affleck-Dine mechanism}

\label{subsec:AD}

The scalar potential vanishes along flat directions in the global SUSY
limit. However, it is lifted by the SUSY-breaking effects, which depend
on the SUSY-breaking mechanism. Moreover it is also lifted by a
nonrenormalizable operator in the superpotential,
\begin{equation}
W=\frac{\Phi^n}{nM^{n-3}},
\end{equation}
where $M$ is a cutoff scale for an interaction and $n$ is an integer,
which depends on a flat direction $\Phi$. For example, $n=6$ if $\Phi$
parameterizes the $\bu \bd \bd$ flat direction
since $(\bu \bd \bd)^2$ is the lowest order nonrenormalizable operator.

In the gravity or anomaly mediated SUSY-breaking scenario, the flat
directions acquire their masses \cite{Enqvist98},
\begin{equation}
V_{\rm grav} \simeq m_{\phi}^2 
\left[1+K\log\left(\dfrac{|\Phi|^2}{M_G^2} \right)\right]|\Phi|^2, \label{gravpot}
\end{equation}
where $m_\phi \sim {\rm TeV}$ is the soft SUSY-breaking mass,
$M_G=1/\sqrt{8\pi G}$ is the reduced Planck mass, and $K$ is a numerical
coefficient coming from one-loop corrections. The sign of $K$ depends on
the details of the loop effects:
$K$ can be positive if the top quark loop effects are the dominant
contribution, which is realized when the top Yukawa coupling is order
unity. On the other hand, $K$ is negative with $K = (-0.01\sim -0.1)$
when the gaugino loop effects are dominant \cite{Enqvist98}.
As shown later, $Q$-balls are formed when $K$ is negative.

In the gauge-mediated SUSY-breaking model, on the other hand, the
potential along the flat direction is lifted as \cite{de Gouvea:1997tn},
\begin{equation}
V_{\rm gauge} \simeq \left\{ 
\begin{array}{ll}
m_{\phi}^2|\Phi|^2 & \quad (|\Phi| \ll M_S), \\
M_F^4 \left(\log \dfrac{|\Phi|^2}{M_S^2} \right)^2
& \quad (|\Phi| \gg M_S), \\
\end{array} \right. \label{gaugepot}
\end{equation}
where $M_F$ is the SUSY-breaking scale and $M_S=M_F^2/m_{\phi}$ is the
messenger mass.  The allowed parameter range of $M_F$ is $10^4{\rm GeV}
\lesssim M_F \lesssim 10^{10}{\rm GeV}$ \cite{de Gouvea:1997tn}.  The
upper bound comes from the condition that the gravity effects should not
be so strong and the lower bound comes from the condition that the SUSY
breaking scale should be larger than the electroweak scale.  The
shape of the potential can be understood by noting that for large
$|\Phi|>M_S$, the supersymmetry-breaking mass terms are suppressed by a
factor of $M_S^2/|\Phi|^2$ and hence the scalar potential becomes
 constant at $|\Phi|>M_S$.
In addition to $V_{\rm gauge}$, there also exists the gravity effects, 
\begin{equation}
V_{\rm grav2} \simeq m_{3/2}^2 \left[1+K\log\left(\dfrac{|\Phi|^2}{M_G^2} \right)\right]|\Phi|^2, \label{gravpot2}
\end{equation}
where $m_{3/2}$ is the gravitino mass ranging from 1 eV to 10 GeV. This
term typically dominates over $V_{\rm gauge}$ at large field values.

The scalar potential also contains the contribution from
nonrenormalizable operators, called $A$-term $V_A$ and $F$-term $V_F$:
\beqa
&&V_A= \frac{a_m m_{3/2}}{M^{n-3}}\Phi^n + {\rm H.c.}, \label{a-term} \\
&&V_F= \frac{|\Phi|^{2n-2}}{M^{2n-6}},  \label{f-term}
\eeqa
where $a_m$ is a complex parameter and its absolute magnitude is less than order unity. 
$A$-term arises from the gravitational interaction between the
nonrenormalizable operator and the SUSY-breaking sector.  

During both inflation
and the inflaton oscillation dominated era, the AD field receives the
Hubble induced mass term coming from the gravitational interaction
between the AD field and the inflaton :
\begin{equation}
V_{\rm HM}=-c_H H^2|\Phi|^2, 
\end{equation}
where $c_H$ is a positive constant of order unity and $H$ is the Hubble parameter.
The balance
between $F$-term and the (negative) Hubble induced mass term determines
the initial value of the AD field, but these terms are irrelevant for the
subsequent dynamics.

In addition to the SUSY-breaking effects and the nonrenormalizable
operators, there are other contributions to the scalar potential. When
the thermal plasma exists, the AD field receives the finite temperature
effects given by \cite{Anisimov},
\begin{equation}
V_{\rm thermal}\sim \left\{
\begin{array}{ll}
h^2 T^2 |\Phi|^2 & \quad (h|\Phi|<T), \\
c \alpha_g^2 T^4\log\left(\dfrac{|\Phi|^2}{T^2}\right) & \quad (h|\Phi|>T), 
\end{array} \right. \label{thermalpot}
\end{equation}
where $h$ is the Yukawa or the gauge coupling constant for the
corresponding AD field, $T$ is the temperature of the thermal plasma,
$c$ is a numerical constant of order unity, and $\alpha_g\equiv
g^2/4\pi$ represents the gauge coupling constant. The sign of $c$
depends on the AD field and we assume it to be positive henceforth.  The
upper term in the right-hand side of Eq. \eqref{thermalpot} represents
the thermal mass from the thermal plasma and the lower one represents
the two-loop finite temperature effects coming from the running of the
gauge coupling $g(T)$ with the nonzero field value of the AD field.
Note that the thermal plasma exists even before the reheating from the inflaton
decay. This is because the partial decay of the inflaton before its
complete decay
generates the thermal plasma as a subdominant component of the Universe.
During the inflaton oscillation dominated era, 
the temperature of the Universe can be expressed as \cite{kolbturner}
\begin{equation}
T\simeq A_T^{1/8}\left(\frac{H M_G T_R^2}{A}\right)^{1/4} (\propto a^{-3/8}),  \label{temposc} 
\end{equation}
where $A \equiv \pi^2 g_*/90$ with $g_*$ being the effective
relativistic degrees of freedom.  The subscript ``$T$'' indicates that
the parameter is evaluated at the scale higher than 1 TeV.  Here we have
assumed that reheating from inflaton decay takes place at the
temperature higher than 1 TeV.  Note that, in the context of MSSM, at
the energy scale above $\sim 1$ TeV, $g_*\simeq 220$, at the energy
scale 100 MeV $\sim$ 1 TeV, $g_*\simeq 100$, at the energy scale 0.1 MeV
$\sim$ 100 MeV, $g_*\simeq 10$ and at the energy scale below $\sim$ 0.1
MeV, $g_*\simeq 4$.  The temperatures at the onset of the AD-field
oscillation, $Q$-ball formation and domination, and reheating, which
are given in Sec.~\ref{subsec:Qfate}, are higher than the
electroweak scale when the amplitude of the GWs is large enough.  Thus,
we neglect the time variation of the relativistic degrees of freedom
before $Q$-ball formation and reheating and we express $A$ at those epochs
as $A_T$ henceforth. On the other hand, the temperatures at $Q$-ball
domination and $Q$-ball decay can be less than 1 TeV and hence we express
$A$ at that time as $A_{\rm dom}$ and $A_{\rm dec}$, respectively.

Now we consider 
the dynamics of the AD field. The total effective potential of the AD
field $V$ is then given by $V=V_{\rm grav/gauge}+(V_{\rm
grav2})+V_A+V_F+V_{\rm thermal}+V_{\rm HM}$, and the AD field obeys the
equation of motion,
\begin{equation}
{\ddot \Phi}+3H{\dot \Phi}+\frac{\partial V}{\partial \Phi^*}=0, 
\end{equation}
where the dot denotes the derivative with respect to the cosmic time,
$t$.  The AD field quickly settles down to the potential minimum $|\Phi|
\simeq (H M^{n-3})^{1/(n-2)}$, which is determined by the balance
between $V_{\rm HM}$ and $V_F$.

$V_{\rm HM}$ decreases after inflation in response to the decrease of
the Hubble parameter and disappears after the reheating of the
Universe. Then, the AD field
(more precisely its radial component) 
begins to oscillate around the origin when
\begin{equation}
H_{\rm osc}^2=V^{\prime \prime}(\Phi), 
\label{hosc}
\end{equation}
where 
the dash denotes the derivative with respect to $\phi\equiv
\sqrt{2}|\Phi|$.
Hereafter the subscript ``osc'' indicates that the parameter or the variable is 
evaluated at the beginning of the oscillation of the AD field. The field
value at which the AD field begins to oscillate is given by $\phi_{\rm osc} \simeq (H_{\rm osc}M^{n-3})^{1/(n-2)}$. 

The potential of the AD field also contains a phase dependent term, that
is, $A$-term, which rotates the AD field unless the phase component of the
AD field, $\theta \equiv \arg [\Phi]$, accidentally sits on the
potential valley of the $A$-term Eq. (\ref{a-term}). As a consequence, the
orbit of the AD field in the phase space becomes elliptical and its
ellipticity is estimated to be $\epsilon \simeq a_m
m_{3/2}/V^{\prime\prime}(\phi_{\rm osc})$.

If the AD field carries baryon or lepton charge $\beta_c$, the angular
momentum of the motion in the complex plane of the AD field represents
the baryon or lepton number density given by
\begin{equation}
n_B(t_{\rm osc})=i\beta_c ({\dot \Phi}^* \Phi-\Phi^*{\dot \Phi})\simeq \beta_c a_m m_{3/2} \phi_{\rm osc}^2, \label{baryon}
\end{equation}
which implies that baryon or lepton asymmetry is generated in the Universe. 

\subsection{$Q$-ball formation}

\label{subsec:Qform}

Next we consider the $Q$-ball formation associated with the AD mechanism.
Fluctuations around the homogeneous mode feel spatial instabilities and
grow nonlinearly during the oscillation of the AD field \cite{Kusenko97}
and eventually form clumpy objects, $Q$-balls, if $V(\phi)/\phi^2$ has a
global minimum at $\phi=\phi_{\rm min} \not = 0$ \cite{Coleman}.  This
is because the pressure of the AD field is negative for such a potential
\cite{Enqvist98}.  From Eqs. \eqref{gravpot}, \eqref{gaugepot}, and
\eqref{thermalpot}, the condition can be realized when $V_{\rm grav(2)}$
with negative $K$, $V_{\rm gauge}(\phi>M_S)$, or $V_{\rm
thermal}(\phi>hT)$ dominates the potential energy.  In fact, it is
confirmed numerically that fluctuations develop and go nonlinear to form
$Q$-balls \cite{kk00a,kk01,kk99}.  Their stabilities are guaranteed by
global $U(1)$ charge, that is, baryon or lepton charge in our case. In
this subsection we briefly review the amplification of the fluctuations
of the AD field and the properties of the subsequently produced
$Q$-balls. Here we concentrate on the case when the dynamics of the AD
field is driven by the thermal logarithmic potential
Eq. \eqref{thermalpot}. In Appendix \ref{App:zero}, we will comment
on the cases when the effective potential is dominated by 
zero-temperature terms (and a negative thermal log term).

First, let us examine the growing of the fluctuations of the AD field
using the linear perturbation analysis. We write the AD field $\Phi$ as
\begin{equation}
\Phi=\frac{\phi}{\sqrt{2}}e^{i\theta}, 
\end{equation}
and decompose the radial $(\phi)$ and the phase $(\theta)$ components
into their homogeneous parts and perturbations,
\begin{align}
\phi(\vecs{x},t)&=\phi(t)+\delta \phi(\vecs{x},t), \\
\theta(\vecs{x},t)&=\theta(t)+\delta \theta(\vecs{x},t). 
\end{align}
Once the baryon or the lepton number is fixed, the phase dependent term
in the potential is irrelevant. Then, neglecting such terms in the
potential, the equations of motion in the flat FRW universe read
\cite{Kusenko97}
\begin{align}
{\ddot \phi}+3H{\dot \phi}-\phi {\dot \theta}^2+V^{\prime}&=0,  \label{EOM1}\\
{\ddot \theta}+3H{\dot \theta}+2\frac{{\dot \phi}}{\phi}{\dot \theta}&=0,  \label{EOM2}\\
\delta {\ddot \phi} +3H\delta{\dot \phi}-\frac{1}{a^2}\Delta \delta \phi-(2{\dot \theta}\phi \delta {\dot \theta}+
  {\dot \theta}^2\delta \phi)+V^{\prime\prime}\delta\phi&=0, \label{EOM3}\\
\delta {\ddot \theta}+3H\delta {\dot \theta}+2\left(\frac{{\dot \theta}}{\phi}\delta {\dot \phi} +
  \frac{{\dot \phi}}{\phi} \delta {\dot \theta}-\frac{\dot\phi\dot\theta}{\phi^2}\delta\phi\right)-\frac{1}{a^2}\Delta \delta \theta 
  &=0.  \label{EOM4}
\end{align}
To find the instability bands, let us
write the perturbations as
\begin{align}
\delta \phi(\vecs{x},t) &=\delta \phi_0e^{S(t)+i{\vecs k}\cdot{\vecs x}}, \\
\delta \theta (\vecs{x},t)&=\delta \theta_0e^{S(t)+i{\vecs k}\cdot{\vecs x}},   
\end{align}
where $\delta \phi_0$ and $\delta \theta_0$ are constants. Inserting these forms into Eqs. \eqref{EOM1}, \eqref{EOM2}, 
\eqref{EOM3}, and \eqref{EOM4}, we have the following dispersion relation, 
\begin{equation}
\left[{\dot S}^2+\left(3H+\frac{2{\dot \phi}}{\phi}\right){\dot S}+\frac{k^2}{a^2}\right]\left[{\dot S}^2 + 3H{\dot S}+\frac{k^2}{a^2}-\omega^2+V^{\prime\prime}\right]+4\omega^2{\dot S}\left({\dot S}-\frac{\dot \phi}{\phi}\right)=0, 
\end{equation}
where $k^2 \equiv |{\vecs k}|^2$ and $\omega \equiv {\dot
\theta}$. The fluctuations with the momentum ${\vecs k}$ grow
exponentially when the condition 
\begin{equation}
\frac{k^2}{a^2}+V^{\prime \prime}-\omega^2<0, 
\end{equation}
is satisfied. Here we have assumed the inflaton oscillation dominated era. The wavenumber of the maximal growth mode is given by
\begin{equation}
\frac{k^2_{\rm max}}{a^2}=\frac{1}{16\omega^2}(7\omega^4 -
6V^{\prime\prime}\omega^2-V^{\prime\prime 2}), \label{growmode}
\end{equation}
and the fastest growing rate $\beta_{\rm gr}$ is given by
\begin{equation}
\beta_{\rm gr}\equiv {\dot S}=\frac{1}{4}\left|\frac{\omega^2-V''}{\omega}\right| <\frac{k_{\rm max}}{a}. 
\label{growrate}
\end{equation}
Finally, such growing fluctuations go nonlinear to form $Q$-balls. 
The Hubble parameter $H_*$ at the $Q$-ball formation may
be expressed as 
\begin{equation}
H_* = \frac{1}{\alpha}\beta_{\rm gr}, \label{h*}
\end{equation}
where $\alpha > 1$ is a numerical factor that represents the duration of the $Q$-ball formation, and $\alpha \simeq \CO(10)$ 
for the case of a mass term with negative $K$ \cite{Kusenko97} 
and $\alpha \simeq \CO(1)$ for the case of a logarithmic term \cite{kk01}. 

Next, we investigate the properties of $Q$-balls. The properties of
$Q$-balls are different for different
types, and here we concentrate on the case when the oscillation of the
AD field is driven by the thermal logarithmic potential,
\begin{equation}
V_{\rm thermal}\simeq \alpha_g^2 T^4\log\left(\dfrac{|\Phi|^2}{T^2}\right)  \label{thermpot}. 
\end{equation}
For the gravity or anomaly mediated SUSY-breaking mechanism, this
situation can realized when
\begin{equation}
\alpha_g^2 T_{\rm osc}^4 \gg m_\phi^2 \phi_{\rm osc}^2,  \label{th_ow_grav}
\end{equation}
and for the gauge-mediated SUSY-breaking mechanism, the thermal
logarithmic potential dominates when
\beqa
\alpha_g^2 T_{\rm osc}^4 \gg \left\{
\begin{array}{lr}
{\rm max} \{M_F^4, m_{3/2}^2 \phi_{\rm osc}^2\} \label{th_ow_gauge1} \ \ &{\rm for} \ \ \phi_{\rm osc}>M_S, \\
 m_{\phi}^2 \phi_{\rm osc}^2 \label{th_ow_gauge2} \ \ &{\rm for} \ \ \phi_{\rm osc}<M_S.  \\
\end{array}
\right. \label{th_ow_gauge}
\eeqa
The angular velocity of the homogeneous Affleck-Dine field,
$\omega$, is determined by the mass scale around $\phi_{\rm osc}$ and is
estimated to be $\sqrt{2} \alpha_g T_{\rm osc}^2/\phi_{\rm osc}$, which
yields from Eq. (\ref{growmode}) and Eq. (\ref{growrate})
\begin{equation}
\frac{k_{\rm max}^2}{a^2}\simeq \frac{3}{2}\frac{\alpha_g^2 T_{\rm osc}^4}{\phi_{\rm osc}^2} \ \ {\rm and} \ \ \beta_{\rm gr} \simeq \frac{\alpha_g T_{\rm osc}^2}{\sqrt{2}\phi_{\rm osc}}.  \label{gmgwthermal}
\end{equation}
The Hubble parameter at the $Q$-ball formation, Eq. (\ref{h*}), becomes
\begin{equation}
H_*\simeq \frac{1}{\alpha} \frac{\alpha_g T_{\rm osc}^2}{\phi_{\rm osc}},   \label{hubform}
\end{equation}
and from Eq. (\ref{temposc}) and Eq. (\ref{hosc}), the temperature at the $Q$-ball formation is
\beqa
T_*\simeq \alpha^{-1/4}T_{\rm osc}.   
\label{t*}
\eeqa
 
Almost all the baryon/lepton charges produced by the AD mechanism are
absorbed into $Q$-balls. Then, the baryon/lepton charge stored in a $Q$-ball
is estimated as
\begin{equation}
Q\simeq  {\bar \epsilon}\beta \left(\dfrac{\phi_{\rm osc}}{\alpha_g^{1/2}T_{\rm osc}}\right)^4, \label{chthermal}
\end{equation}
where ${\bar \epsilon}$ is related to the ellipticity of the orbit of
the AD field, $\epsilon$, as ${\bar \epsilon}=\epsilon$ for $\epsilon >
0.06$ and ${\bar \epsilon}=0.06$ for $\epsilon < 0.06$, and $\beta\simeq
6\times 10^{-4}$ \cite{kk01}.  The numerical factor $\beta$ represents
the dilution due to the cosmic expansion during the $Q$-ball
formation. Here one should notice that the total charge trapped by a
$Q$-ball is proportional to the ellipticity $\epsilon$ for $\epsilon >
0.06$. On the other hand, it saturates for $\epsilon <0.06$, 
since both negatively charged $Q$-balls and positively charged ones
are formed. Other properties of $Q$-balls are given by,
\begin{align}
R&\simeq \frac{Q^{1/4}}{\sqrt{2}\alpha_g^{1/2}T_{\rm osc}}, & \omega&\simeq \frac{\sqrt{2}\pi \alpha_g^{1/2}T_{\rm osc}}{Q^{1/4}}, \notag \\
\phi_Q&\simeq  \alpha_g^{1/2}T_{\rm osc} Q^{1/4},  &E_Q&\simeq \frac{4\pi\sqrt{2}}{3}\alpha_g^{1/2}T_{\rm osc}Q^{3/4},   \label{qthermal}
\end{align}
where $R$ is the radius of $Q$-balls, $\phi_Q$ is the value of $\phi$ at the center of the $Q$-balls, and $E_Q$ is 
the energy per a $Q$-ball. From Eq.~\eqref{qthermal}, the average energy density of $Q$-balls is estimated as 
\begin{equation}
\rho_Q^* \simeq E_Q n_Q \simeq \frac{\alpha_g^2 T_{\rm osc}^4 ({\bar \epsilon}\beta)^{3/4}}{\eta}, \label{rhoq*}
\end{equation}
where $n_Q \simeq (k_{\rm max}/a)^3$ is the number density of $Q$-balls and 
$\eta$ is a numerical factor of order unity. 
{}From Eqs.~\eqref{h*} and \eqref{rhoq*}, the corresponding density parameter is given by
\beqa
\Omega_Q^*=\frac{\rho_Q^*}{3M_G^2H_*^2}\simeq 
\frac{\alpha^2({\bar \epsilon}\beta)^{3/4}}{3\eta}\left(\frac{\phi_{\rm osc}}{M_G}\right)^2.
\label{omegaq*}
\eeqa

\subsection{The fate of $Q$-balls}

\label{subsec:Qfate}

In this subsection, we examine the fate of $Q$-balls, which is important
for estimating the present amplitude of the GWs from the $Q$-ball
formation. Since the dominant contribution to the potential depends on
the temperature which changes with the cosmic time, the properties of
$Q$-balls change as well. Moreover,
for some temperatures, other contributions can dominate the thermal
logarithmic contribution in the potential, which implies that the
properties of $Q$-balls may drastically change and
$Q$-balls may disappear if the dominant contribution of the potential does
not allow a $Q$-ball solution. Finally $Q$-balls can decay from their
surfaces if the energy per charge is smaller than the nucleon mass or
the neutrino mass.  In the following, we investigate the fate of $Q$-balls in detail.

\subsubsection{Transformation of $Q$-balls}

As mentioned above, the potential for the AD field has temperature
dependence and so do the properties of $Q$-balls.  The AD-field value at
the center of the $Q$-balls and the energy per a $Q$-ball are given by
\begin{align}
 \phi_Q(T) &\simeq\alpha_g^{1/2} T Q^{1/4}\simeq ({\bar \epsilon}
 \beta)^{1/4} \phi_{\rm osc}\frac{T}{T_{\rm osc}} \propto T,  \label{thphiq} \\ 
 E_Q(T)&\simeq \frac{4\pi\sqrt{2}}{3}\alpha_g^{1/2}T Q^{3/4}
 \simeq \frac{4\pi\sqrt{2}}{3} \frac{({\bar \epsilon}
 \beta)^{3/4}}{\alpha_g} \left(\frac{\phi_{\rm osc}}{T_{\rm osc}}\right)
 T \propto T. 
\end{align}
Here we have used Eq.~\eqref{chthermal}. 
We can see that $\phi_Q(T)$ and $E_Q(T)$ are proportional to the
temperature of the thermal plasma and hence decreases with the cosmic
time. 
Therefore, the average energy density of $Q$-balls is proportional
to $Ta^{-3}$. 
Since $T\propto a^{-3/8}$ in the inflaton oscillation dominated era, so 
the energy density of $Q$-balls  decreases in proportion to $T a^{-3}\propto  T^9$: 
 \begin{equation}
 \rho_Q\simeq \frac{\alpha^{9/4}\alpha_g^2}{\eta}\frac{T^9}{T_{\rm osc}^5}({\bar \epsilon}\beta)^{3/4}.  \label{rhoqt}
 \end{equation}
After reheating, it decreases in proportion to $Ta^{-3}\propto T^4$.  
Thus, as long as $V_{\rm thermal}$ dominates the potential of $Q$-balls, $Q$-balls never
dominate the energy density of the Universe.\\

\paragraph{Gravity or anomaly mediated SUSY-breaking model \\ \\}

\begin{figure}[htbp]
 \begin{center}
  \includegraphics[width=100mm]{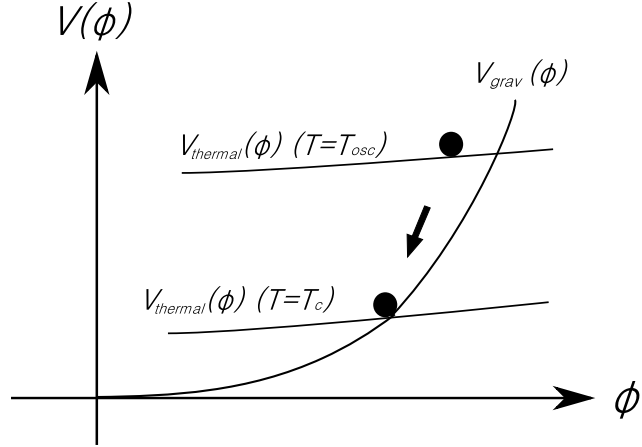}
 \end{center}
 \caption{The scalar potential $V_{\rm grav}+V_{\rm thermal}$ is shown. The horizontal axis is 
the amplitude of the AD field $\phi=\sqrt{2}|\Phi|$. The filled circle represents the position of 
the AD field. Initially its rotation is driven by $V_{\rm thermal}$ at $T=T_{\rm osc}(>T_c)$. 
At the temperature $T=T_c$, $V_{\rm grav}$ dominates over $V_{\rm thermal}$, and 
the properties of $Q$-balls are changed. }
 \label{fig:Qthgrav}
\end{figure}

While $V_{\rm thermal}$ is proportional to $T^4$, $V_{\rm grav}\simeq
\dfrac{1}{2}m_\phi^2\phi_Q^2(T)$ is proportional to $T^2$. Thus, in the
case of the gravity or anomaly mediated SUSY-breaking model, using
Eq.~\eqref{thphiq}, $V_{\rm grav}$ dominates $V_{\rm thermal}$ at the
temperature below $T_c$ given by
\begin{equation}
T_c=\frac{({\bar \epsilon}\beta)^{1/4}}{\alpha_g}m_\phi \frac{\phi_{\rm osc}}{T_{\rm osc}}.  
\label{tc}
\end{equation}
Once $V_{\rm grav}$ becomes dominant in the potential of the AD field,
the properties of $Q$-balls are changed.  Figure \ref{fig:Qthgrav} shows
the schematic form of this transition. As the temperature decreases,
$V_{\rm thermal}$ and $\phi_Q$ also decrease. At the temperature $T_c$,
the dominant contribution to the potential is changed.
In the gravity or anomaly mediated SUSY-breaking model with negative
$K$, the potential still allows $Q$-balls but their properties are changed
as follows:
\begin{align}
R^2&\simeq \frac{2}{m_{\phi}^2|K|}, & \omega&\simeq m_{\phi}, \notag \\
\phi_Q&\simeq \left(\frac{|K|}{\pi}\right)^{3/4}m_{\phi}Q^{1/2}, & E_Q &
 \simeq \frac14 m_{\phi} Q. 
 \label {qgrav}
\end{align}
Here, we have used the fact that $Q$ is conserved through the
transition. One should notice that such $Q$-balls behave like matter
because $E_Q$ does not decrease after the transition. In the gravity or
anomaly mediated SUSY-breaking model with positive $K$, on the other
hand, the $Q$-ball solution no longer exists, which makes $Q$-balls
unstable. Therefore the AD field would be almost homogeneous. The energy
density of such an AD field also behaves like matter. However, there is
an important difference between the cases with positive $K$ and negative
$K$. While $Q$-balls can decay only from their surfaces,
the almost homogeneous AD field can decay over the whole space. Thus,
$Q$-balls can survive longer and dilute the produced GWs further.\\

\paragraph{Gauge-mediated SUSY-breaking model \\ \\}

In the gauge-mediated SUSY-breaking model, the situations are rather
complicated. While $V_{\rm thermal}$ is proportional to $T^4$ and
$V_{\rm grav2}$ is proportional to $T^2$, $V_{\rm gauge}$ is independent
of $T$. Thus, there are three possibilities depending on which term dominates the 
potential next. 

\begin{itemize}

\item {\it Case A: $V_{\rm gauge}$ driven $Q$-ball transformation. }

One is that $V_{\rm gauge}$ dominates $V_{\rm thermal}$ at the critical
temperature given by
\begin{equation}
T_c^A=\alpha_g^{-1/2}M_F, \label{tca}
\end{equation}
so that the type of the $Q$-ball is changed into the gauge mediation type. 
The properties of this type of $Q$-balls are given by
\begin{align}
R&\simeq \frac{Q^{1/4}}{\sqrt{2}M_F}, & \omega&\simeq \frac{\sqrt{2}\pi M_F}{Q^{1/4}}, \notag \\
\phi_Q&\simeq  M_F Q^{1/4},  &E_Q&\simeq \frac{4\pi\sqrt{2}}{3}M_FQ^{3/4} .  \label{qthgauge}
\end{align}
Note the fact that $Q$ is conserved through the
transition.  In this case $V_{\rm grav2}$ never becomes the dominant
contribution in the potential of the AD field.

\item {\it Case B : $V_{\rm grav} (K < 0) $ driven $Q$-ball transformation. }

Another is that $V_{\rm grav2}$ with negative $K$ dominates 
at the critical temperature given by
\begin{equation}
T_c^B=\frac{({\bar \epsilon}\beta)^{1/4}}{\alpha_g}m_{3/2} \frac{\phi_{\rm osc}}{T_{\rm osc}},  \label{critemnew}
\end{equation}
so that the type of the $Q$-ball is changed into the new type.  The
properties of this type of $Q$-balls are given by
\begin{align}
R^2&\simeq \frac{2}{m_{3/2}^2|K|}, & \omega&\simeq m_{3/2}, \notag \\
\phi_Q&\simeq \left(\frac{|K|}{\pi}\right)^{3/4}m_{3/2}Q^{1/2}, & E_Q &
 \simeq \frac14 m_{3/2} Q. 
 \label {qnew}
\end{align}
Here, $Q$ is also conserved at the transition. In this case, $V_{\rm
gauge}$ never becomes the dominant contribution in the potential of the
AD field.

\item {\it Case C : $V_{\rm grav} (K>0) $ driven $Q$-ball transformation. }

The other one is that $V_{\rm grav2}$ with positive $K$
dominates first at the critical temperature $T_c^B$ given in Eq.
\eqref{critemnew}. This potential does not allow a $Q$-ball solution and
hence the almost homogeneous AD field is recovered. After this
transition, the AD field decreases as $\phi\propto H$ because of the
cosmic expansion. Thus, $V_{\rm gauge}$ dominates 
$V_{\rm grav2}$ at
$\phi=\phi_{eq} \equiv M_F^2/m_{3/2}$, which implies that $Q$-balls are
formed again. The properties of this type of $Q$-balls are those
of the delayed type $Q$-balls and are given by
\begin{align}
R&\simeq \frac{Q^{1/4}}{\sqrt{2}M_F} \simeq (\sqrt{2}m_{3/2})^{-1}, & \omega&\simeq \frac{\sqrt{2}\pi M_F}{Q^{1/4}}\simeq \sqrt{2} \pi m_{3/2}, \notag \\
\phi_Q&\simeq  \alpha_g^{1/4} M_F Q^{1/4} \simeq \frac{M_F^2}{m_{3/2}},  &E_Q&\simeq \frac{4\pi\sqrt{2}}{3}M_FQ^{3/4} .  \label{qthdelay}
\end{align}
In this case, $Q$ is given by\begin{equation}
Q\simeq \left(\frac{\phi_{eq}}{M_F}\right)^4 \simeq \left(\frac{M_F}{m_{3/2}}\right)^4.
\end{equation}

\end{itemize}

\begin{figure}[htbp]
\begin{tabular}{cc}
\begin{minipage}{0.5\hsize}
\begin{center}
\includegraphics[width=80mm]{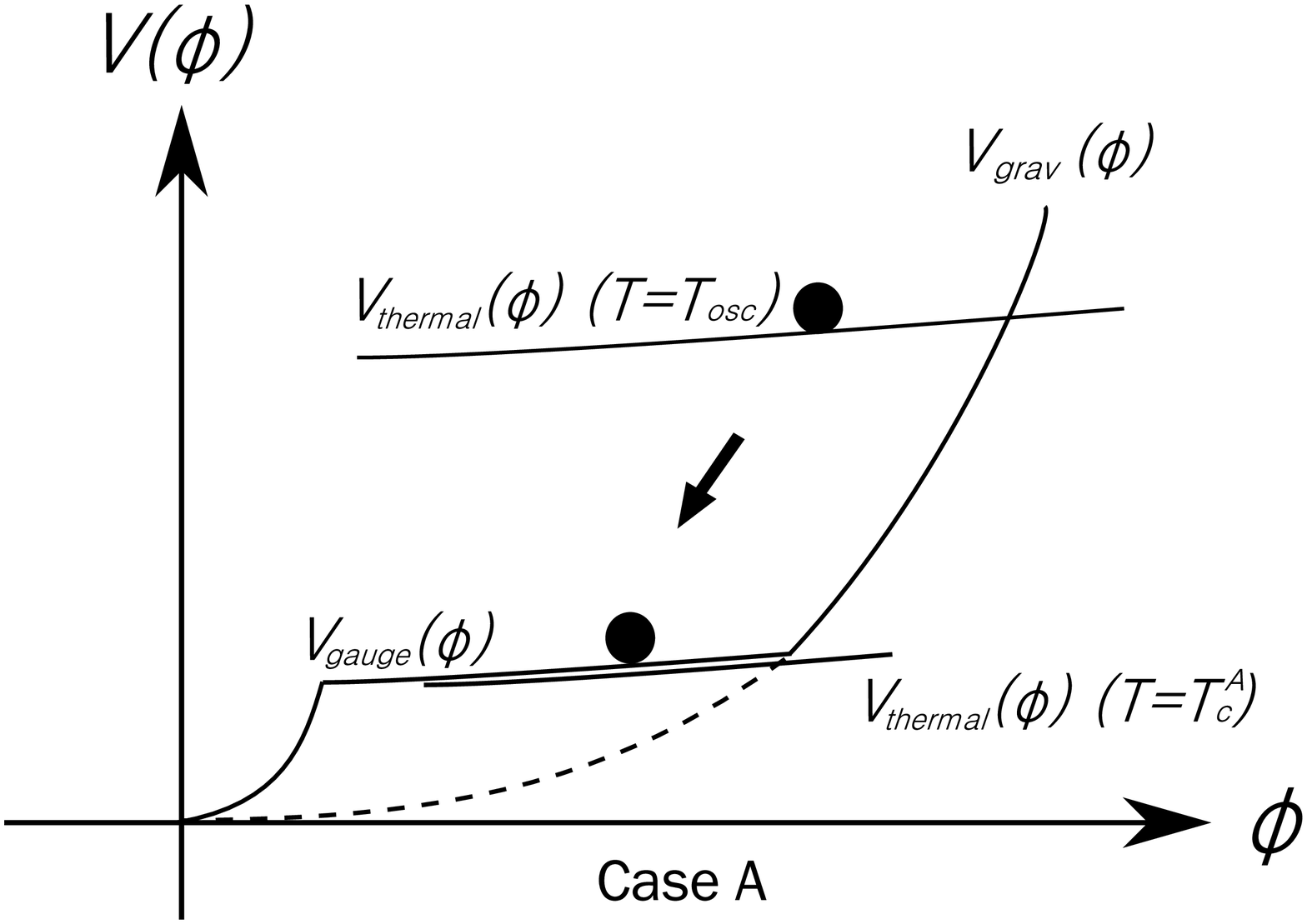}\end{center}
\end{minipage}
&
\begin{minipage}{0.5\hsize}
\begin{center}
\includegraphics[width=80mm]{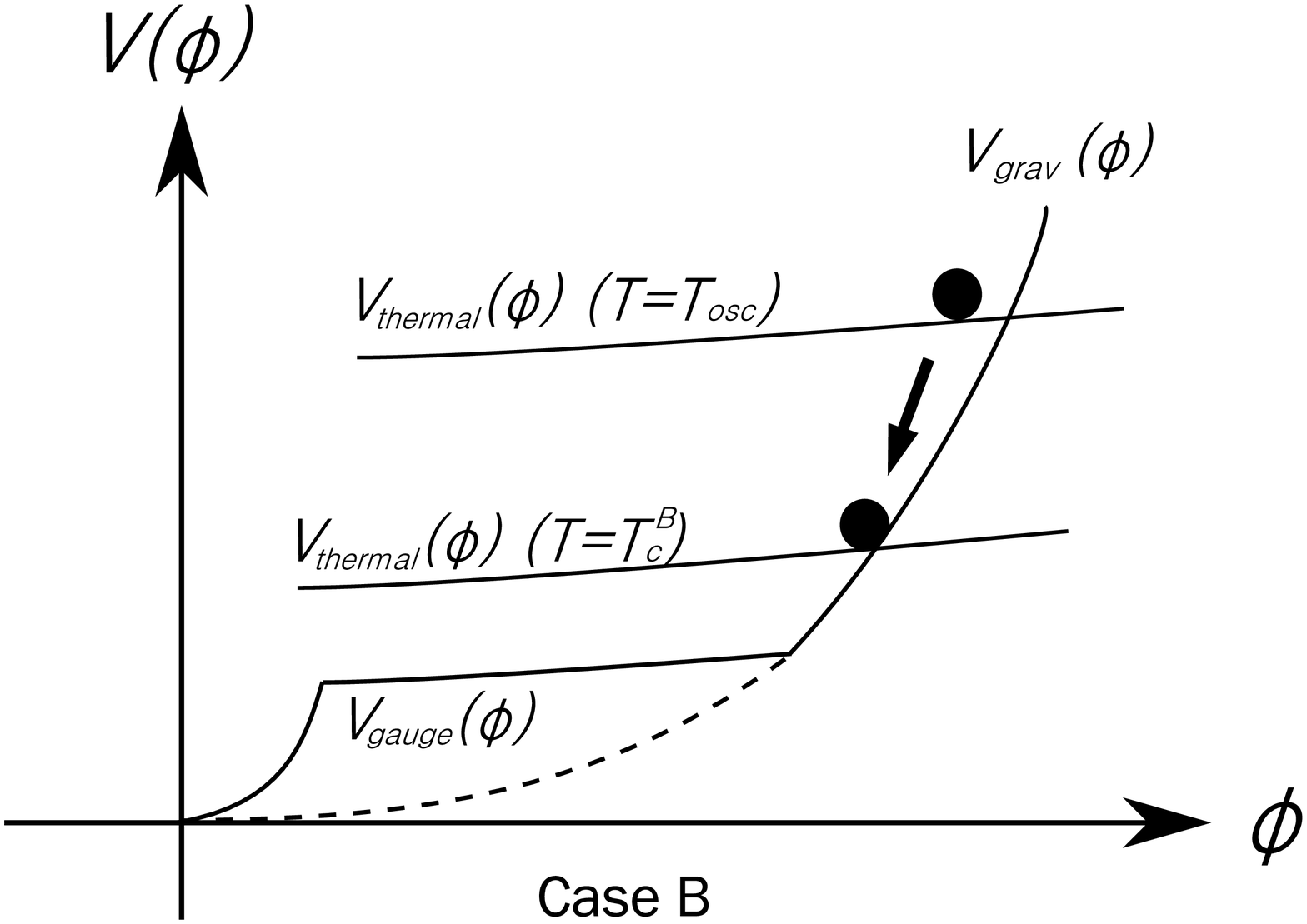}
\end{center}
\end{minipage}
\\
\begin{minipage}{0.5\hsize}
\begin{center}
\includegraphics[width=80mm]{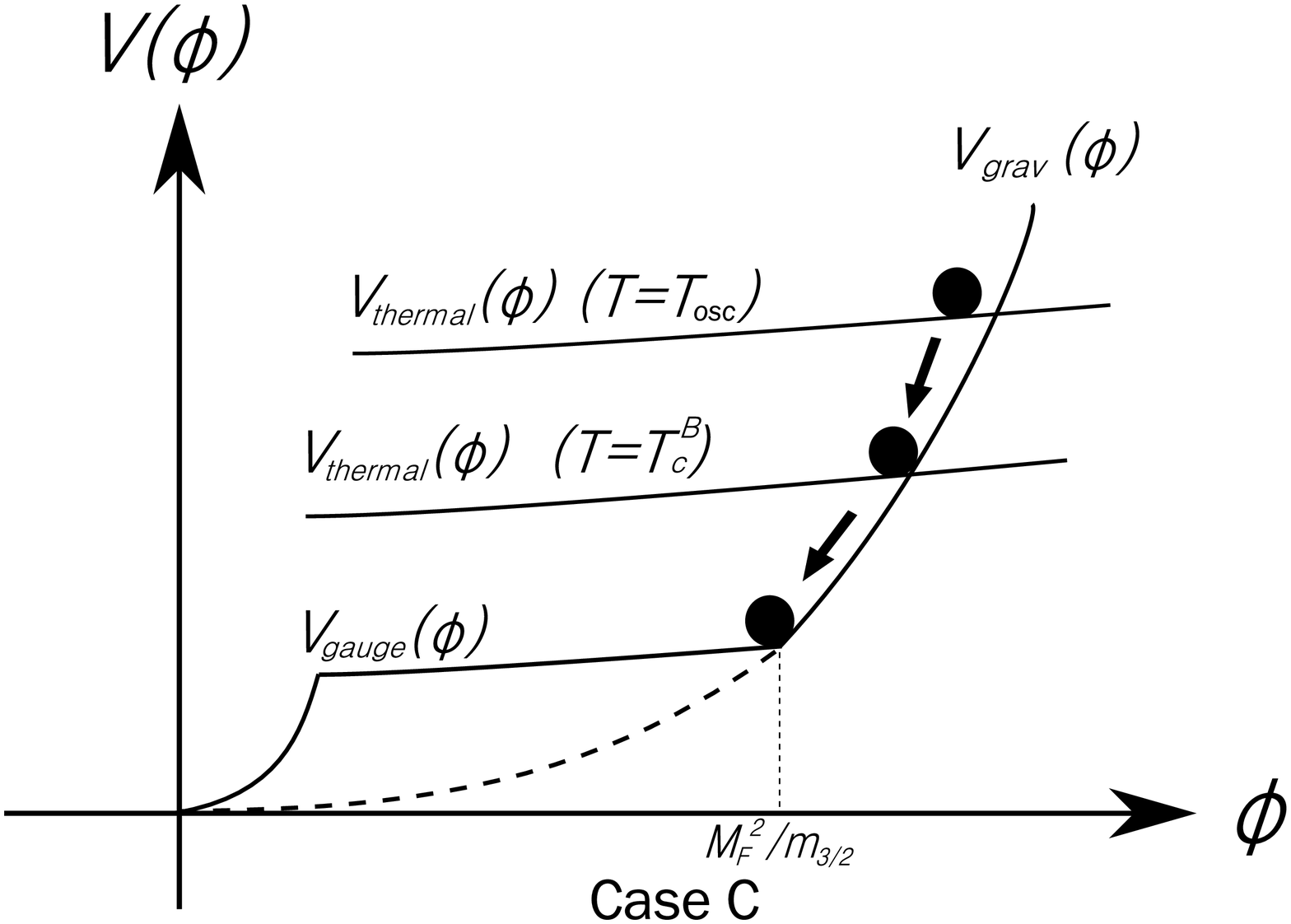}
\end{center}
\end{minipage}
\end{tabular}
\caption{The scalar potential in the gauge-mediated SUSY-breaking model in each case is shown. 
The horizontal axis is the amplitude of the AD field. The filled circle represents 
the position of the AD field. Initially its rotation is driven by $V_{\rm thermal}$ at 
$T=T_{\rm osc}$ and dominant potential terms change at $T=T_c^{A(B)}$. There are 
three possibilities depending on which term dominates afterwards. }
\label{fig:gauge}
\end{figure}

In all cases, $Q$-balls behave like matter because $E_Q$ does not decrease
after the transition and hence $Q$-balls may dominate the energy density
of the Universe. Figure \ref {fig:gauge} shows the schematic form of the
transition of each case. As the temperature decreases, $V_{\rm thermal}$
and $\phi_Q$ also decrease. The dominant contribution to the potential
is changed at the temperature $T_c^{A(B)}$, and hence the type of
$Q$-balls are changed as well. 

\subsubsection{Decay of $Q$-balls and AD field}

$Q$-balls can decay into light fermions if the decay processes are
kinematically allowed. However, in their interiors the Pauli exclusion
principle forbids their decays into fermions \cite{Cohen:1986ct}. 
Therefore $Q$-balls can decay only from their surfaces. This sets the upper bound on the decay
rate of $Q$-balls,
\begin{equation}
\left|\frac{dQ}{dt}\right| \leq \frac{\omega^3R^2}{48 \pi}.  \label{decrate}
\end{equation}
In fact, it is almost saturated for the cases we are interested in \cite{Cohen:1986ct}. 
Then, we can express the decay rate of $Q$-balls as 
\begin{equation}
\Gamma_{\rm dec}\equiv \frac{1}{Q}\frac{dQ}{dt}=\frac{\omega^3 R^2}{48 \pi Q}, \label{decgen} 
\end{equation}
and the Hubble parameter at the $Q$-ball decay is given by $H_{\rm
dec}=\Gamma_{\rm dec}$\footnote{If there are light bosons and the AD
field can decay into them by two-loop interactions, the decay width
would be enhanced by a factor $f_s \lesssim 10^3$. However, such bosons
do not exist in the context of MSSM because scalar fields other
than the flat direction acquire large masses in general, and hence we
do not consider such decay processes. }.  The decay rate of the $Q$-balls
is different according to their types.  In the case of the gravity or
anomaly mediated SUSY-breaking models with negative $K$, from
Eq.~\eqref{qgrav}, the decay rate of $Q$-balls can be expressed as
\begin{equation}
\Gamma_{\rm dec}\equiv \frac{1}{Q}\frac{dQ}{dt} \simeq \frac{1}{Q} \frac{m_\phi}{24 \pi |K|}. \label{hdecneg} 
\end{equation}
As mentioned above, in the gauge-mediated SUSY-breaking case, there are
three possibilities for the fate of $Q$-balls.
{}From Eqs.~\eqref{qthgauge}, \eqref{qnew}, \eqref{qthdelay}, and \eqref{decrate}, the decay rate can be expressed as
\begin{align}
\Gamma_{\rm dec}&\equiv \frac{1}{Q}\frac{dQ}{dt} \simeq \left\{
\begin{array}{ll}
\dfrac{1}{Q^{5/4}} \dfrac{\pi^2 M_F}{24\sqrt{2} }  \ \ &{\rm for \ Case \ A}, \\
\dfrac{1}{Q} \dfrac{m_{3/2}}{24 \pi |K|}   \ \ &{\rm for \ Case \ B}, \\
\dfrac{\pi^2 m_{3/2}^5}{24\sqrt{2}M_F^4}  \ \ &{\rm for \ Case \ C}. 
\end{array}\right. \label{decga}
\end{align}
The decay temperature should be higher than $1$~MeV for the successful
big bang nucleosynthesis (BBN), which constrains the parameters of
$Q$-balls, such as $\phi_{\rm osc}$ and $T_R$, if $Q$-balls dominate the
energy density of the Universe.\footnote{More precisely, if $Q$-balls
contribute to more than about $1$\% of the energy density of the
Universe, their decay temperature should be higher than $1$~MeV for the
successful BBN. But, this constraint is not so important for our
estimate, and we ignore such small difference.}  The constraint in each
case will be given in the next section.

Note that $Q$-balls can evaporate when there are thermal plasmas, as
discussed in Ref. \cite{Laine:1998rg}. Though $Q$-balls can evaporate away
before their decays, this takes place only when $Q$ is small enough,
which is unfavorable for our purpose because the initial amplitudes of
the produced GWs would also be small. In Appendix A, we will give 
more quantitative discussion on the charge evaporation from $Q$-balls.

On the other hand, in the case of the gravity or anomaly mediated SUSY
breaking models with positive $K$, $Q$-balls vanish at $T=T_c$ and hence
the almost homogeneous AD field is recovered, whose amplitude
decreases as $\phi\propto H$ due to the cosmic expansion. As long as $h
\phi_Q> m_\phi$, the fields coupled with the AD field acquire large
masses so that the AD field cannot decay into them. 
When $h \phi_Q \simeq m_\phi$,
the decay into the light fermions is allowed kinematically. The decay
rate is given by
\begin{equation}
\Gamma_{\rm dec}\simeq\frac{h^2}{8\pi} m_\phi. 
\end{equation}
In the case where the AD field dominates the Universe at that time, the
Hubble parameter is given by
\begin{equation}
H_{\rm dec} = \frac{m_\phi^2}{\sqrt{6}h M_G}. \label{hdecpos}
\end{equation}
Therefore, if $h\gtrsim (m_\phi/M_G)^{1/3}\simeq 10^{-5}$, then, $\Gamma_{\rm dec} 
\gtrsim H_{\rm dec}$, and the AD field decays 
into light fermions quickly  at that time. The decay temperature $T_{\rm dec}$ is estimated as 
\begin{equation}
T_{\rm dec}\simeq \frac{m_\phi}{(6A_T)^{1/4}h^{1/2}}>m_\phi. 
\end{equation}
Thus, the AD field can decay before the BBN and the electroweak symmetry breaking.

Before closing this section, we comment on the decay rate of $Q$-balls
adopted in Ref. \cite{Kusenko:2008zm}. The authors of
Ref. \cite{Kusenko:2008zm} considered two possibilities of the $Q$-ball
decay to enhance its decay rate. One is to introduce higher-dimensional
operators, which preserve the SUSY but violate the baryon
numbers. However, if such operators provide the decays into fermions,
the decay rate is strongly constrained by the Pauli blocking as
mentioned before. We also would like to point out that it is difficult
for such operators to provide the decays into bosons in the context of
MSSM, because such bosons acquires large masses through the Yukawa
couplings so that the decays into them are kinematically
prohibited. Furthermore, even if we find such bosons,
the decay rate estimated in Ref.  \cite{Kusenko:2008zm} cannot be
applied to $Q$-balls, as pointed out by the authors themselves.  The other
is the semiclassical decay through the $A$-terms, which was investigated
in detail in Ref.  \cite{Kawasaki:2005xc}. In Ref.
\cite{Kawasaki:2005xc},
it is found that the instability due to the $A$-term causes the $Q$-ball
decay in the Minkowski background. Then the authors of
Ref. \cite{Kusenko:2008zm} simply apply the expression of the decay
rate to the case of the expanding Universe, which is not justified, and
conclude that $Q$-balls quickly decay via this process. In the expanding
Universe, however, this process is not effective because the amplitude
of the AD field decreases and the $A$-term becomes soon negligible
due to the cosmic expansion. In fact, the authors of
Ref. \cite{Kawasaki:2005xc} conclude that $Q$-balls cannot decay via this
process in the expanding Universe. Therefore, in this paper, we use the
conservative estimate of the decay rate of $Q$-balls, different from
Ref. \cite{Kusenko:2008zm}.

\section{Gravitational Waves from $Q$-balls}

\label{sec:GWs}

In this section we investigate the features of the GWs produced from the
$Q$-ball formation and discuss the prospect for the detection of such GWs.
First we estimate the amplitudes and the typical frequencies of the GWs
at the $Q$-ball formation. Then, we evaluate the dilution of the GWs
during the subsequent cosmic expansion.  As we have seen in the
Sec. \ref{subsec:Qfate}, $Q$-balls driven by the thermal logarithmic
potential 
never dominate the energy density of the Universe until their
transformation. For this reason, we concentrate on the situation where
$Q$-balls are formed when the thermal logarithmic potential is the
dominant contribution to the potential of the AD field.  The other cases
where $Q$-balls are produced for the zero-temperature potential are
discussed in Appendix B.

\subsection{Generation of GWs}

\label{subsec:genGW}

In this subsection, we study the generation of the GWs associated with
the fragmentation of the AD field and estimate the amplitudes and
frequencies of the GWs.  The process of the fragmentation of the AD
field is inhomogeneous and nonspherical so that GWs may be emitted at
the $Q$-ball formation.

We evaluate the initial amplitudes and frequencies of the GWs
{}from the $Q$-ball formation, by using the equations for the
transverse-traceless (TT) component of the metric perturbations,
following Refs. \cite{Kusenko:2008zm,garcia-bellido}. By perturbing the
Einstein equation, we obtain the equation for the TT component of the
metric perturbations,
\begin{equation}
{\ddot h_{ij}^{\rm TT}}({\vecs x},t)+3H{\dot h_{ij}^{\rm TT}}({\vecs x},t)-\frac{\nabla^2}{a^2}h_{ij}^{\rm TT}({\vecs x},t)
=16 \pi G \Pi_{ij}^{\rm TT}({\vecs x},t),  \label{evo1}
\end{equation}
where $h_{ij}^{\rm TT}({\vecs x},t)$ is the TT component of the metric
perturbation and $\Pi_{ij}^{\rm TT}({\vecs x},t)$ is the TT component of the
energy-momentum tensor of the AD field. 
Instead of using the above Eq. \eqref{evo1}, 
it is easier to use the
following equations: 
\begin{equation}
{\ddot u_{ij}}({\vecs k},t)+3H{\dot u_{ij}}({\vecs k},t)+\frac{{\vecs k}^2}{a^2}u_{ij}({\vecs k},t) =16 \pi G T_{ij}({\vecs k},t).   \label{evo2}
\end{equation} 
Here $u_{ij}({\vecs k},t)$ and $T_{ij}({\vecs k},t)$ satisfies the relations 
\begin{align}
h_{ij}^{\rm TT}({\vecs k},t) &= \Lambda_{ij,mn}(\hat{\vecs k}) u_{mn}({\vecs k},t), \\
\Pi_{ij}^{\rm TT}({\vecs k},t) &=\Lambda_{ij,mn}(\hat{\vecs k}) T_{mn}({\vecs k},t), \\
T_{ij}({\vecs x},t)& = \frac{1}{a^2} \partial_i \phi({\vecs x},t) \partial_j \phi({\vecs x},t),
\end{align}
where the projection tensor $\Lambda_{ij,mn}$ is defined by 
\begin{align}
\Lambda_{ij,mn}(\hat{\vecs k})&\equiv \left(P_{im}(\hat{\vecs k})P_{jn}(\hat{\vecs k})-\frac{1}{2}P_{ij}(\hat{\vecs k})P_{mn}(\hat{\vecs k})\right), \label{proj} \\
P_{ij}(\hat{\vecs k})&\equiv \delta_{ij}-{\hat k}_i{\hat k}_j, 
\end{align}
with ${\hat k}_i \equiv k_i/|{\vecs k}|$.  $h_{ij}^{\rm TT}({\vecs
k},t)$ and $\Pi_{ij}^{\rm TT}({\vecs k},t)$ are Fourier transforms of
$h_{ij}^{\rm TT}({\vecs x},t)$ and $\Pi_{ij}^{\rm TT}({\vecs x},t)$,
respectively.  Since Eq. \eqref{evo2} contains unphysical (gauge)
degrees of freedom, we need to follow the time evolution of the variable
$h_{ij}^{\rm TT}$ instead of $u_{ij}$, strictly speaking.
However, in the absence of spherical symmetry and homogeneity, it is
sufficient to approximate $h_{ij}^{\rm TT}$ by $u_{ij}$
\cite{Kusenko:2008zm}.

The energy density of the GWs is given by 
\begin{equation}
\rho_{\rm GW}=\frac{1}{32\pi G L^3}\int  d^3{\vecs x} {\dot h}_{ij}^{\rm TT}({\vecs x},t) {\dot h}_{ij}^{\rm TT}({\vecs x},t) \label{rhogw1}, 
\end{equation}
which can be approximated as 
\begin{equation}
\rho_{\rm GW}\simeq\frac{1}{32\pi G L^3}\int  d^3{\vecs x} {\dot u}_{ij}({\vecs x},t) 
{\dot u}_{ij}({\vecs x},t) \label{rhogw2},  
\end{equation}
where $V=L^3$ is the volume of the space, and we have used the fact that
the process of fragmentation is nonspherical and $|h_{ij}^{\rm TT}|
\simeq |u_{ij}|$ in this situation.

We now estimate the energy density of the GWs produced by the
fragmentation of the AD condensate.  For this purpose, we first evaluate
the energy-momentum tensor of the AD field $T_{ij}$.  The maximal growth
mode of the fluctuation of the AD condensate $\delta \phi$ evolves as
\begin{equation}
\delta \phi ({\vecs x},t) =\delta \phi_0 e^{\beta_{\rm gr} t+ i {\vecs k_{\rm max}}\cdot {\vecs x}},  \label{delphit}
\end{equation}
where $\delta \phi_0$ is the initial value of the field perturbation.
Then, $T_{ij}$ can be estimated as $T_{ij}\simeq -k_{\rm max}^2\delta
\phi^2/3a^2$ after angle average, and Eq. \eqref{evo2} becomes
\begin{equation}
{\ddot u_{ij}}(2{\vecs k}_{\rm max}, t)+\frac{4 k_{\rm max}^2}{a^2} u_{ij}(2{\vecs k}_{\rm max}, t) \simeq
-\frac{2}{3M_G^2}\frac{k_{\rm max}^2}{a^2}\delta \phi_0^2 e^{2\beta_{\rm gr} t},  
\label{evo3}
\end{equation}
where 
the Hubble friction term is neglected since $k_{\rm max}/a$ is larger
than $H$.  Since $\beta_{\rm gr}$ is smaller than $k_{\rm max}/a$
(Eq. (\ref{growrate})), we may use the second term in the left-hand side
in Eq. (\ref{evo3}) to estimate $u_{ij}$ to give 
\begin{equation}
u_{ij}\simeq - \frac{1}{6M_G^2}\delta \phi^2. \label{uij} 
\end{equation}
After a time interval $\Delta t \simeq \ln(\phi_Q/\delta
\phi_0)/\beta_{\rm gr}$, the fluctuations of the flat direction reach
$\delta \phi \simeq \phi_Q$, which implies that $Q$-balls are formed.
Therefore, at the last stage of the formation of $Q$-balls, $\rho_{\rm
GW}$ reaches the maximal value, which is estimated using
Eqs. (\ref{rhogw2}), \eqref{delphit} and \eqref{uij} as
\begin{equation}
\rho_{\rm GW}\simeq \frac{M_G^2}{4}\dot u_{ij}\dot u_{ij}\simeq 
\frac{\beta_{\rm gr}^2}{9M_G^2}\phi_Q^4. 
\label{rhogw}
\end{equation}
The typical frequency of the GWs from the $Q$-ball formation is given by
the wave number of the maximal growing mode, $f_*\simeq k_{\rm max}/(\pi
a)$.  The density parameter of the GWs, $\Omega_{\rm GW}(f_*)$, is then given by 
\begin{equation}
 \Omega_{\rm GW}^*(f_*)\simeq \frac{\beta_{\rm gr}^2\phi_Q^4}{27M_G^4 H_*^2}, \label{omggw*}
\end{equation}
where $H_*$ is the Hubble parameter at the $Q$-ball formation. The present
density parameter and the frequency of the GWs from $Q$-balls are then
given by \beqa &&\Omega_{\rm GW}^0=\Omega_{\rm
GW}^*\left(\frac{a_*}{a_0}\right)^4\left(\frac{H_*}{H_0}\right)^2,\\
&&f_0=f_*\left(\frac{a_*}{a_0}\right).  \eeqa Therefore, in addition to
the redshift factor, the present density parameter contains the dilution
factor due to the matter/$Q$-balls domination effects.

In the case when the oscillation of the AD field is driven by the
thermal logarithmic potential, from Eqs. \eqref{gmgwthermal},
\eqref{hubform}, \eqref{chthermal} and \eqref{qthermal}, the
density parameter and the frequency of the GWs at the $Q$-ball formation are given by
\beqa
&&\Omega_{\rm GW}^* = \frac{\alpha^2}{54}({\bar \epsilon}\beta)
\left(\frac{\phi_{\rm osc}}{M_G}\right)^4,  \label{th_omgwf} \\
&&f_*= \frac{\sqrt{6}\alpha_g}{2\pi}\frac{T_{\rm osc}^2}{\phi_{\rm osc}}. \label{th_f}  
\eeqa

It may be instructive to give alternative derivation of the
energy density of the GWs using the quadrupole approximation. By taking
the moment of inertia as $I \simeq E_QR^2 \simeq \phi_Q^2
\omega^{2}(k_{\rm max}/a)^{-5}$ and approximating the time derivative by
$\beta_{\rm gr}\sim k_{\rm max}/a\sim \omega$, the quadrupole
approximation gives the luminosity $L_{\rm GW} \simeq {\dddot
I}^2M_G^{-2}$ and thus energy density liberated in GWs during the $Q$-ball
formation (the interval $\Delta t\simeq \beta_{\rm gr}^{-1}$) as
\beqa
\rho_{\rm GW}\sim \frac{1}{\beta_{\rm gr}M_G^{2}}\left(\frac{\phi_Q^2 \beta_{\rm gr}^3 \omega^2 a^5}{k_{\rm max}^5} \right)^2n_Q
\sim \frac{\beta_{\rm gr}^2}{M_G^2}\phi_Q^4,
\label{rhogw:q-appx}
\eeqa
where we have used \eqref{gmgwthermal} and \eqref{qthermal} and 
$n_Q\simeq (k_{\rm max}/a)^3 \sim \beta_{\rm gr}^3$. Apart from the numerical factor, 
Eq. (\ref{rhogw:q-appx}) nicely coincides with Eq. (\ref{rhogw}). 

\subsection{Cosmic history}

\begin{figure}[htbp]
\begin{tabular}{cc}
\begin{minipage}{0.5\hsize}
\begin{center}
\includegraphics[width=80mm]{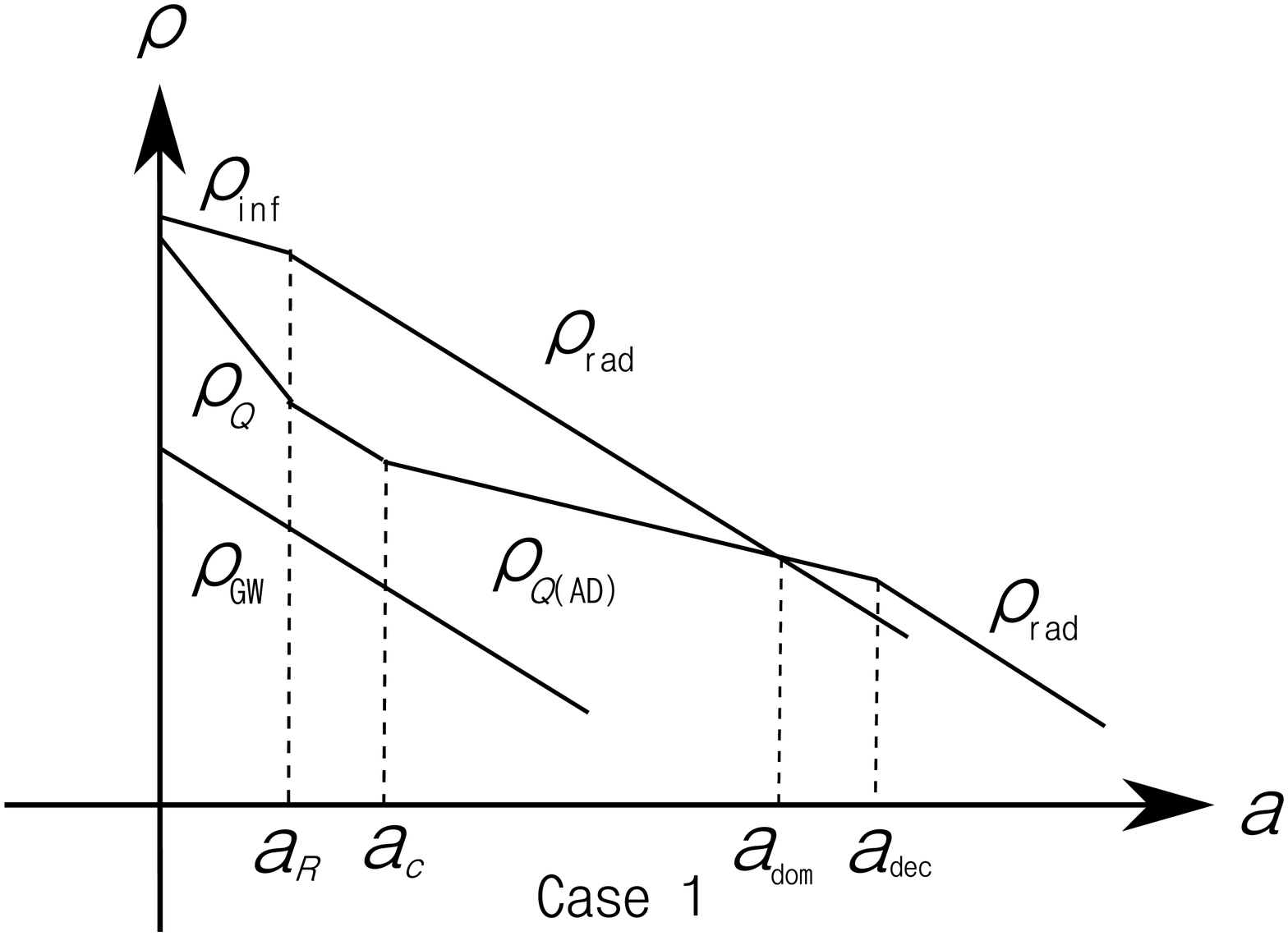}\end{center}
\end{minipage}
&
\begin{minipage}{0.5\hsize}
\begin{center}
\includegraphics[width=80mm]{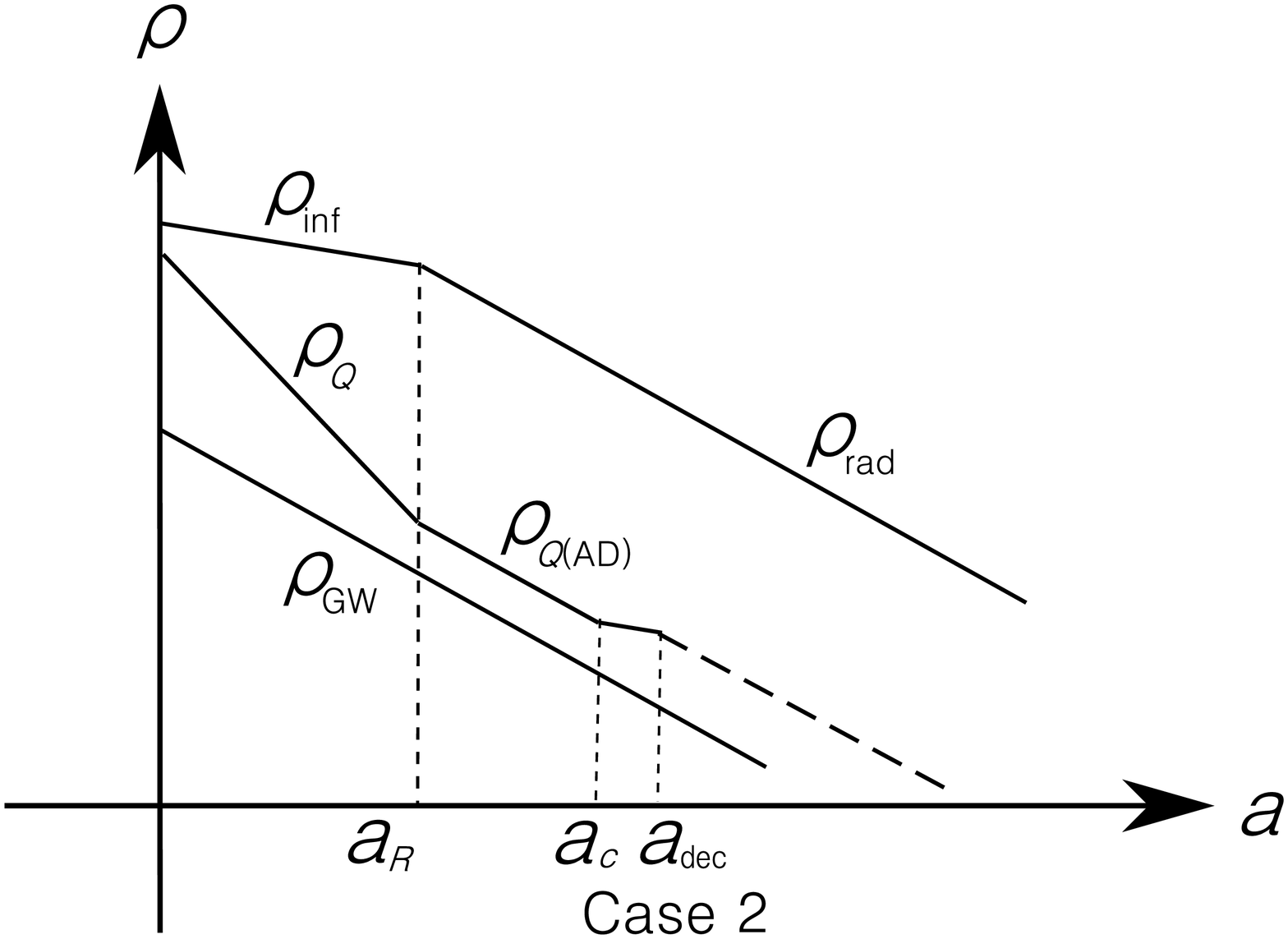}
\end{center}
\end{minipage}
\\
\begin{minipage}{0.5\hsize}
\begin{center}
\includegraphics[width=80mm]{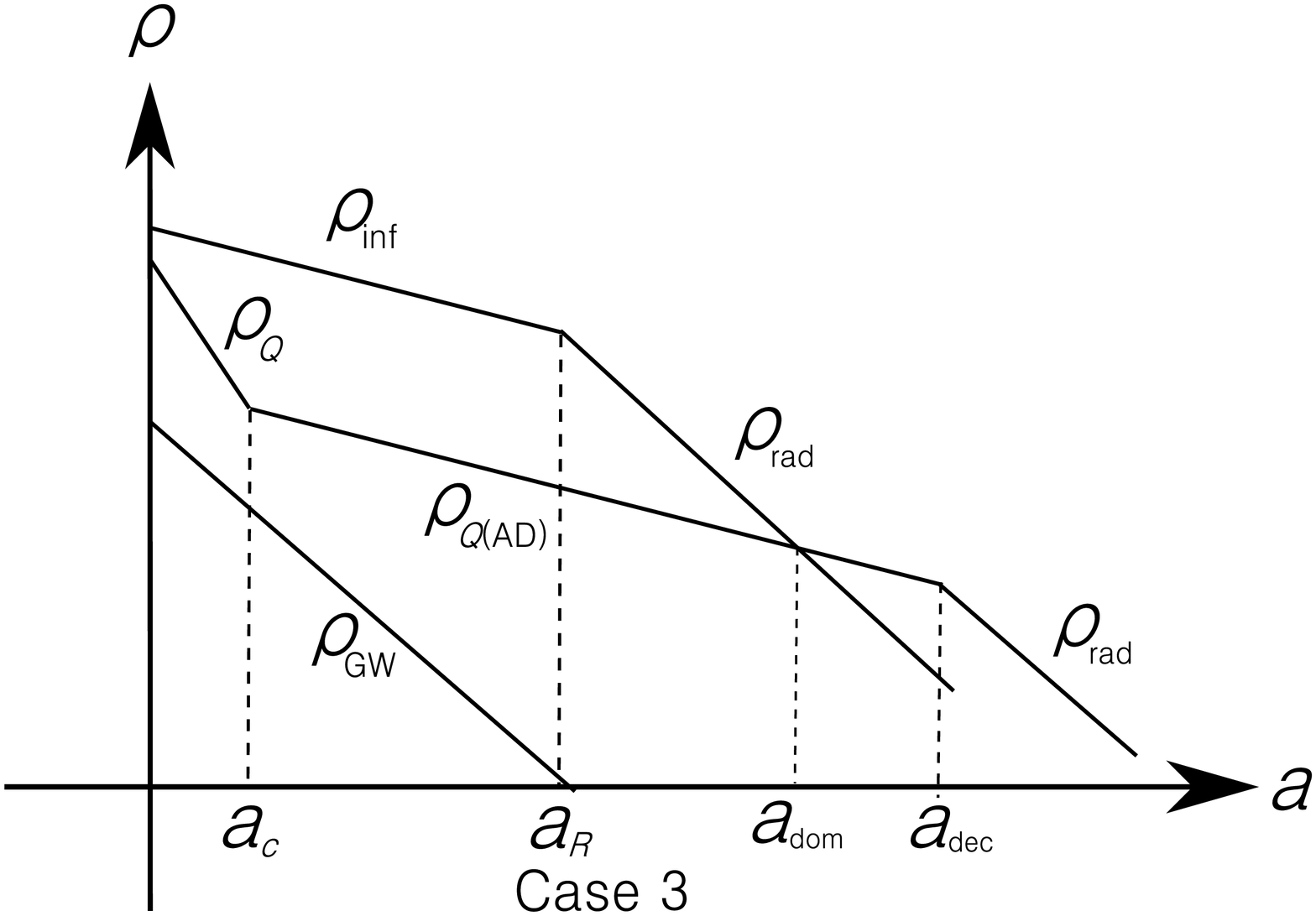}
\end{center}
\end{minipage}
&
\begin{minipage}{0.5\hsize}
\begin{center}
\includegraphics[width=80mm]{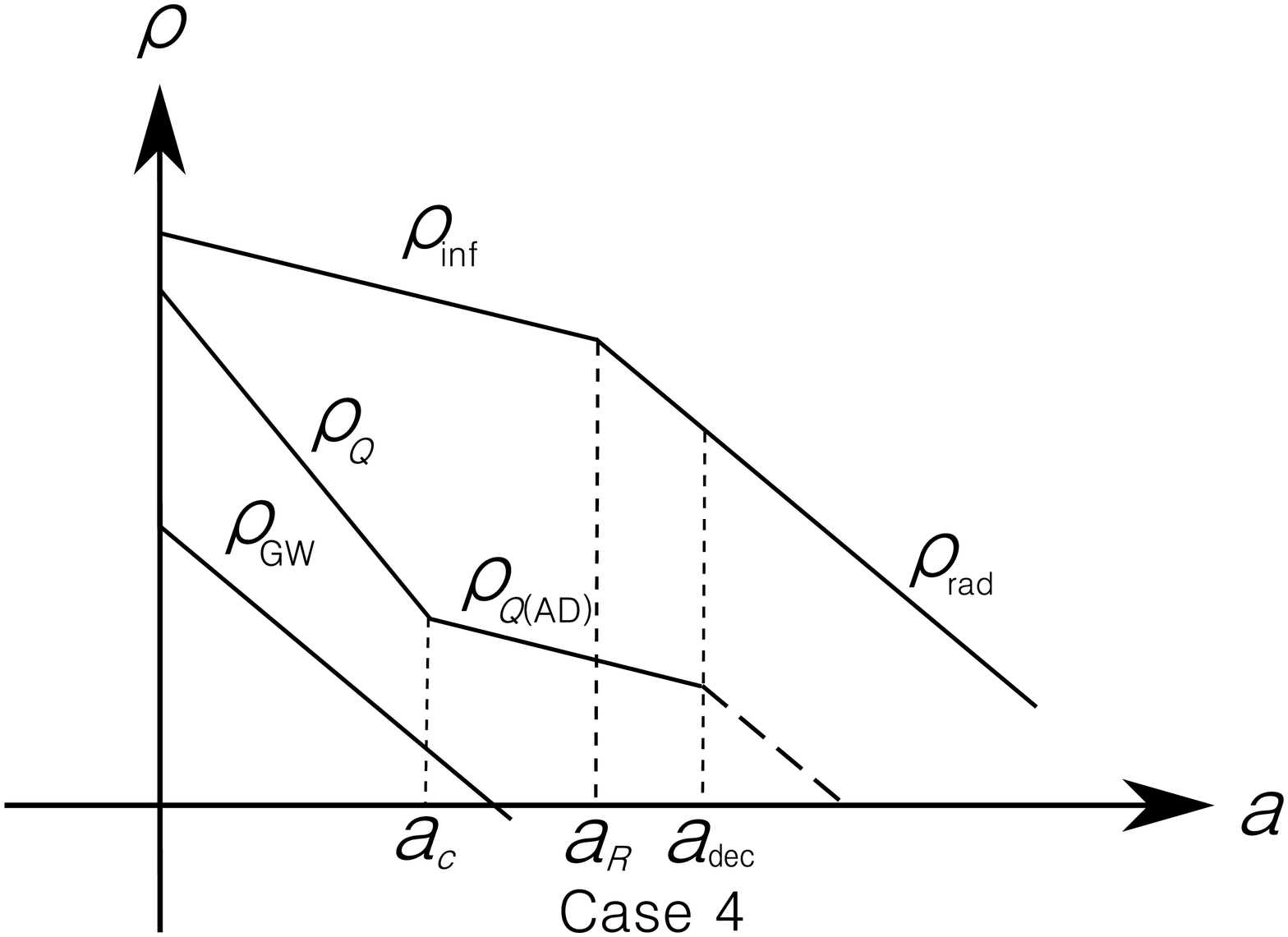}
\label{fig:}
\end{center}
\end{minipage}
\end{tabular}
\caption{The schematic time evolution of the energy density of each
component (inflaton, radiations from the inflaton decay, $Q$-balls,
radiations from the $Q$-ball decay, GWs) for each case in Table
\ref{Table:2} is shown. The vertical axis represents the energy density 
and the horizontal axis represents the scale factor $a$. The dilution
factor of the GWs from the $Q$-ball formation is different in each case.
Even during the inflaton oscillation dominated era, the energy density
of the thermal log type $Q$-balls changes as the temperature of the
thermal plasma generated by partial inflaton decay
decreases. \label{Fig:posikene} }
\end{figure}

In order to evaluate the present amount and typical frequency of the GWs
{}from the $Q$-ball formation, we need to take into account of the cosmic history 
 after the $Q$-ball formation.  In the case where $Q$-balls are
formed when the thermal logarithmic term is the dominant contribution to
the potential of the AD field, there are four possibilities of the
cosmic history. They are classified according to the following two
conditions. The first condition is whether the $Q$-ball dominated era
exists or not, and it is characterized by $H_{\rm dom}$ and $H_{\rm
dec}$.  $H_{\rm dom}$ is the Hubble parameter when the $Q$-balls (or the
AD field) would dominate the energy density of the Universe if such an
epoch would exist.  $H_{\rm dec}$ is the Hubble parameter when the
$Q$-ball (or the AD field) decays.  The second condition is whether 
the reheating after the inflaton decay occurs before the transformation of
the $Q$-balls due to the change of the effective potential or it occurs
after. This condition is characterized by $T_R$ and $T_c$. Here $T_R$ is
the reheating temperature of the inflaton and $T_c$ is temperature at
which the zero-temperature potential dominates over the thermal
logarithmic potential and hence the properties of the $Q$-balls are
changed.  The four cases are shown in Table \ref{Table:1}. 
The time evolution of the energy density of each component of the
Universe in each case is shown in Fig. \ref{Fig:posikene}. The dilution
factor of the GWs from $Q$-ball formation is different in each case.

\begin{table}[htbp]
\begin{center}
\begin{tabular}{l|c}\hline 
Case 1 & Reheating  $\Rightarrow$ $Q$-ball transformation $\Rightarrow$ $Q$-ball/AD-field domination  $\Rightarrow$ $Q$-ball/AD-field decay \\ \hline
Case 2 & Reheating   $\Rightarrow$ $Q$-ball transformation$\Rightarrow$ $Q$-ball/AD-field decay (No $Q$-ball/AD-field domination)\\
\hline
Case 3& $Q$-ball transformation $\Rightarrow$ Reheating $\Rightarrow$ $Q$-ball/AD-field domination  $\Rightarrow$ $Q$-ball/AD-field decay\\ \hline
Case 4 & $Q$-ball transformation $\Rightarrow$ Reheating $\Rightarrow$ $Q$-ball/AD-field decay (No $Q$-ball/AD-field domination) \\ \hline
\end{tabular}
\caption{The possible cosmic histories. \label{Table:1}} 
\end{center}
\end{table}

We note that the following relations for $H_*$ and $T_{\rm osc}$, which
are derived using Eqs. (\ref{temposc}), (\ref{hosc}), (\ref{thermpot})
and (\ref{hubform}), are useful in the subsequent discussion,
\begin{align}
H_* &\simeq \frac{\alpha_g^2 }{\alpha A_T^{1/2}}\frac{T_R^2M_G}{\phi_{\rm osc}^2}& 
&{\rm and}& T_{\rm osc}&\simeq \frac{1}{A_T^{1/4}}\left(\frac{\alpha_g M_G}
{\phi_{\rm osc}}\right)^{1/2}T_R.  
\label{parawithtr} 
\end{align}

\subsection{Gravity or anomaly mediated SUSY-breaking model}

\label{subsec:detectgw} 

In this subsection, we evaluate the present amount and the typical
frequency of the GWs from the $Q$-ball formation in the case where the
$Q$-balls are formed when the thermal logarithmic term is the dominant
contribution to the potential of the AD field in the gravity or anomaly mediated SUSY-breaking model. 
In this case, from Eqs. \eqref{th_ow_grav} and \eqref{parawithtr}, we
have the lower bound on the reheating temperature
\begin{equation}
T_R>\frac{A_T^{1/4}}{\alpha_g}\left(\frac{m_\phi}{M_G}\right)^{1/2}\phi_{\rm osc}\equiv T_R^{c}.     \label{th_ow_grav2}  
\end{equation}
This bound strongly constrains the frequencies of the produced GWs.

\subsubsection{Gravity or anomaly mediated SUSY-breaking model with positive $K$}
\label{subsubsec:posk}

First, we consider the gravity or anomaly mediated SUSY-breaking model
with positive $K$. In this case, when the zero-temperature potential dominates the
effective potential, the $Q$-balls become unstable and hence decay.  Therefore the
almost homogeneous AD field is recovered at the critical temperature
$T_c$, Eq. (\ref{tc}), as discussed in Sec. III C.

First of all, we investigate the condition $T_R>T_c$ in detail, which
discriminates Cases 1 and 2 ($T_R>T_c$) from Cases 3 and 4 ($T_R<T_c$).  Inserting the
expression of $T_{\rm osc}$ Eq. \eqref{parawithtr} into Eq. \eqref{tc}
yields
\begin{equation}
T_c=\frac{A_T^{1/4}}{\alpha_g^{3/2}}({\bar \epsilon}\beta)^{1/4}\frac{m_\phi \phi_{\rm osc}^{3/2}}{M_G^{1/2}T_R}, \label{tcposk}
\end{equation}
which shows that $T_R>T_c$ is equivalent to
\beq
T_R> \frac{A_T^{1/8}}{\alpha_g^{3/4}}({\bar \epsilon}\beta)^{1/8}\frac{m_\phi^{1/2}\phi_{\rm osc}^{3/4}}{M_G^{1/4}}\equiv T_R^{c1}. 
\eeq
Next, in order to examine the condition of the existence of 
the $Q$-ball dominated era, we evaluate the Hubble parameter at the
would-be $Q$-ball domination, $H_{\rm dom}$. 
In fact, $H_{\rm dom}$ in Case 1 and that of Case 3  coincide with each other. 
This can be understood as follows. 
$\Omega_Q \propto T$ during the inflaton oscillation dominated era with thermal log type 
$Q$-balls and $\Omega_Q \propto T^{-1}$ during the radiation dominated era with nonthermal type $Q$-balls, 
while $\Omega_Q$ is constant both during the inflaton oscillation dominated era with zero-temperature type 
$Q$-balls and during the radiation dominated era with thermal log type $Q$-balls. 
Thus, both for Case 1 and Case 3, the following relation is satisfied: 
\begin{equation}
\Omega_Q^{\rm dom}=\frac{T_R T_c}{T_* T_{\rm dom}}\Omega_{Q}^*=1. 
\end{equation}
Therefore we have 
\begin{equation}
H_{\rm dom}=\frac{A_{\rm dom}^{1/2}}{M_G}\frac{T_R^2 T_c^2}{T_*^2}\Omega_Q^{*2}, \label{hdomgene}
\end{equation}
and from Eqs. \eqref{t*}, \eqref{omegaq*}, \eqref{parawithtr} and \eqref{tcposk}, we obtain 
\begin{equation}
H_{\rm dom}=\frac{\alpha^{9/2}A_{\rm dom}^{1/2} A_T}{9\alpha_g^4 \eta^2} 
({\bar \epsilon}\beta)^{2}\dfrac{m_\phi^2\phi_{\rm osc}^8}{M_G^7T_R^2}. \label{hdomposk}
\end{equation}
Here the subscript ``dom'' indicates that the parameter or variable is
evaluated at the would-be $Q$-ball domination.  On the other hand,
$H_{\rm dec}$ is given by Eq. \eqref{hdecpos}.  Comparing these
equations, the condition of the existence of the $Q$-ball dominated era is
equivalent to
\begin{equation}
T_R<\frac{6^{1/4}\alpha^{9/4}h^{1/2} A_{\rm dom}^{1/4} A_T^{1/2}}{3 \eta \alpha_g^2}({\bar \epsilon}\beta)\frac{\phi_{\rm osc}^4}{M_G^3} \equiv T_R^{c2}. 
\end{equation}
{}From $T_c$ in Eq. \eqref{tcposk}, $H_{\rm dom}$ in Eq. \eqref{hdomposk}, and
$H_{\rm dec}$ in Eq. \eqref{hdecpos}, we thus obtain the following
conditions on $T_R$ for each cosmic history, which
are summarized in Table \ref{Table:2}. 

\begin{table}[htbp]
\begin{center}
\begin{tabular}{l|l}
\hline
Case 1 & \ $T_R^{c1}(\phi_{\rm osc})<T_R<T_R^{c2}(\phi_{\rm osc})$  \\ \hline
Case 2 & \ $T_R>{\rm max} \{T_R^{c1}(\phi_{\rm osc}), T_R^{c2}(\phi_{\rm osc})\}$\\
\hline
Case 3 & \ $T_R<{\rm min} \{T_R^{c1}(\phi_{\rm osc}), T_R^{c2}(\phi_{\rm osc})\}$  \\ \hline
Case 4 & \ $T_R^{c2}(\phi_{\rm osc})<T_R<T_R^{c1}(\phi_{\rm osc})$\\ \hline
\end{tabular} 
\end{center}
\caption{The conditions of four cases of the cosmic history. \label{Table:2}}
\end{table}

\begin{figure}[htbp]
 \begin{center}
  \includegraphics[width=150mm]{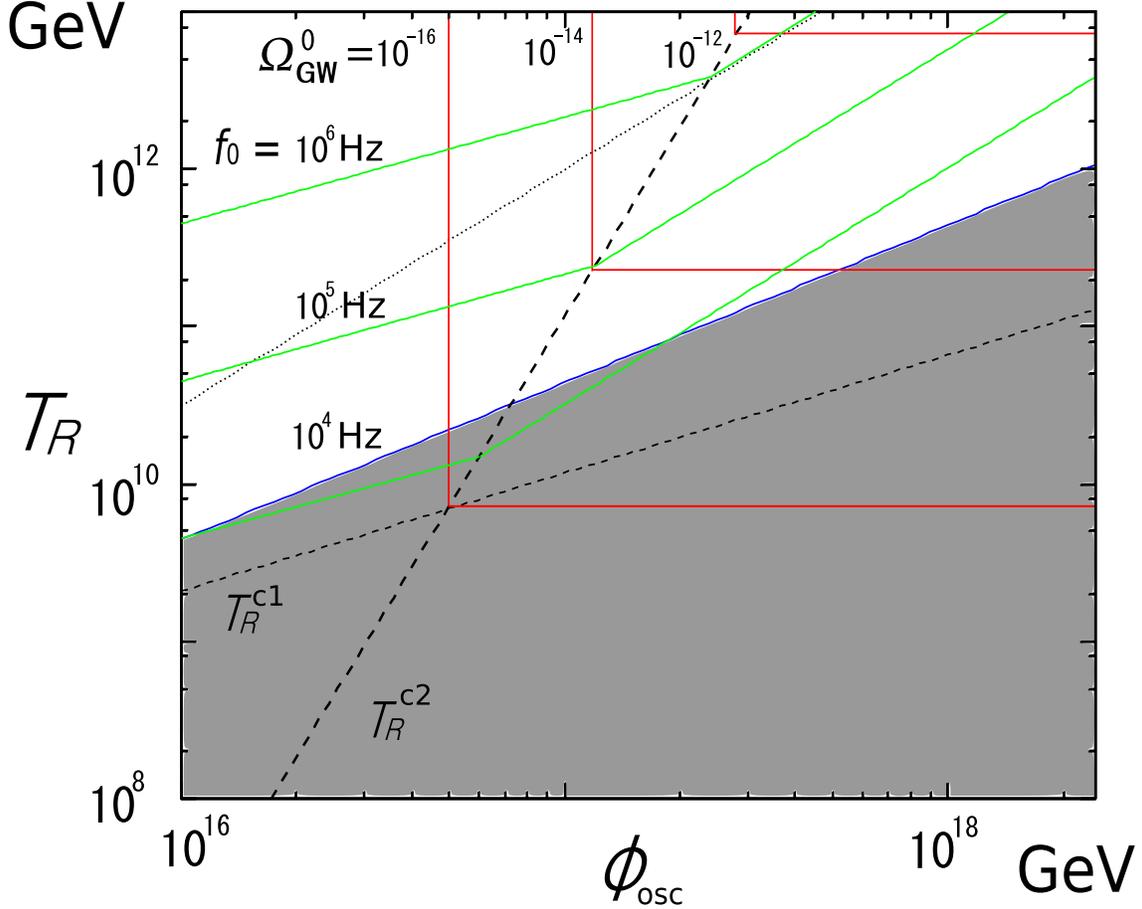}
 \end{center}
 \caption{The contour plot of the present density parameter of the GWs,
$\Omega_{\rm GW}^0=10^{-12}, 10^{-14}$ and $10^{-16}$ (red lines) and its
frequency, $f_0=10^4 $ Hz, $10^5$ Hz, and $10^6$ Hz (green lines) is
shown.  Dashed lines represent $T_R^{c1}$ and $T_R^{c2}$, which
distinguish Cases 1-4 (Table \ref{Table:2}).  In the region surrounded
by $T=T_R^{c2}$ and $\phi_{\rm osc}=M_G$, there is the AD-field
dominated era.  Dotted line represents the boundary of the evaporation
of $Q$-balls, Eq. (\ref{phievap}). In the shaded region, thermal log
type $Q$-balls are not formed, $T_R<T_R^c$. The parameters are taken to
be $ \eta \simeq 1, h\simeq 0.1, \alpha \simeq 1$, $\alpha_g\simeq 0.1$,
and $m_\phi\simeq 10^3$ GeV, and the effective relativistic degrees of
freedom of MSSM are assumed for $A_T, A_{\rm dom}, A_{\rm dec}$ and
$A_0$.}  \label{fig:contour1}
\end{figure}

We then calculate the present density parameters of the GWs, $\Omega_{\rm
GW}^0$, and their frequencies, $f_0$, for each case.  
\begin{align}
\Omega_{GW}^0&=D \Omega_{GW}^*, \\
D&=\left\{
\begin{array}{ll}
\dfrac{a_*}{a_R}\dfrac{a_{\rm dom}}{a_{\rm dec}}\dfrac{a_{eq}}{a_0}  & \quad {\rm (Case \ 1,3)}, \\ 
\dfrac{a_*}{a_R}\dfrac{a_{eq}}{a_0}  & \quad {\rm (Case \ 2,4)},
\end{array} \right. \notag \\
&= \left\{
\begin{array}{ll}
\left(\dfrac{H_R}{H_*}\right)^{2/3}\left(\dfrac{H_{\rm dec}}{H_{\rm dom}}\right)^{2/3}\dfrac{a_{eq}}{a_0} & \quad {\rm (Case \ 1,3)}, \\ 
\left(\dfrac{H_R}{H_*}\right)^{2/3}\dfrac{a_{eq}}{a_0} & \quad {\rm (Case \ 2,4)}. 
\end{array} \right.  \label{dilute}
\end{align}
Here $D$ represents the dilution factor of the GWs due to the matter/$Q$-balls domination. 
The redshift at the $Q$-ball formation is given by
\begin{align}
\frac{a_*}{a_0}&=\left\{
\begin{array}{ll}
\left(\dfrac{A_0}{A_T}\right)^{1/3} \left(\dfrac{A_{\rm dom}}{A_{\rm dec}}\right)^{1/12} 
\dfrac{T_0}{T_R}\left(\dfrac{H_{\rm dec}}{H_{\rm dom}}\right)^{1/6}
\left(\dfrac{H_R}{H_*}\right)^{2/3} & \quad {\rm (Case \ 1,3)}, \\ 
\left(\dfrac{A_0}{A_T}\right)^{1/3} \dfrac{T_0}{T_R}\left(\dfrac{H_R}{H_*}\right)^{2/3} & \quad {\rm (Case \ 2,4)},  
\end{array} \right. \label{redshift}
\end{align}
where the subscript ``0'' indicates that the parameter is evaluated at
present.  Here we have neglected the difference between $g_*$ and
$g_{*s}$, where $g_{*s}$ is the relativistic degrees of freedom for the
entropy density.  $a_{eq}$ and $a_0$ are the scale factors at the
matter-radiation equality and at present, respectively, and so that
$a_0/a_{eq}\simeq 3200$.  From Eqs.~\eqref{hdecpos},
\eqref{th_omgwf}, \eqref{th_f}, \eqref{parawithtr}, \eqref{hdomposk},
\eqref{dilute}, and \eqref{redshift}, we thus obtain the present
density parameter of the GWs, $\Omega_{\rm GW}^0$, as
\begin{align}
\Omega_{\rm GW}^0&\simeq\left\{
\begin{array}{ll}
\dfrac{1}{18\cdot 2^{1/3}}\dfrac{\eta^{4/3}\alpha_g^{4/3}}{\alpha^{1/3} 
A_{\rm dom}^{1/3}h^{2/3}}({\bar \epsilon}\beta)^{-1/3} \left(\dfrac{T_R}{M_G}\right)^{4/3} &\dfrac{a_{eq}}{a_0} \quad {\rm (Case \ 1, 3)},  \\ 
\dfrac{\alpha^{8/3}A_T^{2/3}}{54\alpha_g^{4/3}} ({\bar \epsilon}\beta) 
\left(\dfrac{\phi_{\rm osc}}{M_G}\right)^{16/3}\dfrac{a_{eq}}{a_0} & \quad {\rm (Case \ 2,4)},
\end{array} \right.
\end{align}
 and the present frequency, $f_0$, as
\begin{align}
f_0&\simeq\left\{
\begin{array}{ll}
\dfrac{3^{3/4}}{2^{7/12}\pi}\dfrac{\eta^{1/3}\alpha_g^{4/3}A_0^{1/3}}{\alpha^{1/12}  A_{\rm dec}^{1/12}A_T^{1/3}h^{1/6}} ({\bar \epsilon}\beta)^{-1/3}
\dfrac{M_G^{2/3}T_R^{4/3}}{\phi_{\rm osc}^2}T_0  & \quad {\rm (Case \ 1,3)}, \\ 
\dfrac{\sqrt{6}\alpha_g^{2/3}\alpha^{2/3}A_0^{1/3}}{2 \pi A_T^{1/6}} \dfrac{T_R}{\phi_{\rm osc}^{2/3}M_G^{1/3}}T_0 & \quad {\rm (Case \ 2,4)}.
\end{array} \right.
\end{align}
In Fig. \ref{fig:contour1} we show the contour plot of the present
density parameter of the GWs, $\Omega_{\rm GW}^0$, and their frequency,
$f_0$.  Although the amplitudes of the GWs can be large, their typical
frequencies ($\gtrsim 10^{3 \sim 4}$ Hz at least) are too high to be
detected by the future experiments, since the sensitivity ranges of
future detectors are $\Omega_{\rm GW}^0\simg 10^{-7}$ at $f_0 \simeq
10^{2\sim3}$ Hz for advanced LIGO, $\Omega_{\rm GW}^0\simg 10^{-11}$ at
$f_0 \simeq 10^{-3 \sim -2}$ Hz for LISA, and $\Omega_{\rm GW}^0\simg
10^{-16}$ at $f_0 \simeq 10^{-1}\sim 10$ Hz for DECIGO.  Large
$\Omega_{\rm GW}^0$ requires high reheating temperature and high decay
rate, but that results in higher frequency of the GWs in turn.

In the above argument, we do not consider the possibility that the
$Q$-balls evaporate out via diffusion processes before their decay.
However, that could be possible only when charge $Q$ stored in a $Q$-ball
is very small.  This can be understood as follows. First one should
notice that from Eqs. \eqref{chthermal}, \eqref{parawithtr}, and
\eqref{tcposk}, the typical charge of the $Q$-ball and the transformation
temperature are given by
\begin{equation} 
Q \simeq 10^{-4}\frac{\phi_{\rm osc}^6}{M_G^2T_R^4}  \ \ {\rm and} \ \ T_c \simeq\frac{m_\phi \phi_{\rm osc}^{3/2}}{M_G^{1/2}T_R}.  \label{thq}
\end{equation}
Inserting these values into Eq. \eqref{evthq1} in Appendix A yields 
\begin{equation}
\phi_{\rm osc} \lesssim 10^{8/15}\frac{T_R^{2/3}M_G^{7/15}}{m_\phi^{2/15}} 
\simeq 10^{15}{\rm GeV} \left(\frac{T_R}{10^9 {\rm GeV}}\right)^{2/3} 
\left(\frac{m_\phi}{1 {\rm TeV}}\right)^{-2/15} \equiv \phi_{\rm osc}^{\rm evap}. 
\label{phievap}
\end{equation}
In Fig. \ref{fig:contour1}, this constraint is represented by dotted
line  above which $Q$-balls evaporate out before their transformation. 
{}From Fig, \ref{fig:contour1}, we find that the evaporation is effective at 
$f_0\gtrsim 10^6$ Hz  for $\Omega_{\rm GW}^0>10^{-16}$.
Therefore the charge evaporation from the $Q$-balls does not change our conclusion.

\subsubsection{Gravity or anomaly mediated SUSY-breaking model with negative $K$}

\label{subsubsec:negk}

Next we discuss 
the gravity or anomaly mediated SUSY-breaking model with negative $K$.
The only difference between the cases with positive and negative $K$ is
that $Q$-balls
do not disappear even after the critical temperature $T_c$ for negative
$K$.  Thus, the cosmic history can be classified to four possibilities
in the same way as the case with positive $K$. However, the decay time
of the $Q$-balls, $H_{\rm dec}$, is different from that of the AD field
for positive $K$. From Eqs. \eqref{chthermal}, \eqref{hdecneg}, and
\eqref{parawithtr}, it can be estimated as
\begin{equation}
H_{\rm dec}=\frac{\alpha_g^4  }{24\pi|K| A_T}({\bar \epsilon}\beta)^{-1}
\frac{M_G^2 T_R^4}{\phi_{\rm osc}^6} m_\phi. \label{dechnegk}
\end{equation}
In contrast to the case with positive $K$, the decay temperature 
($A_{\rm dec}^{-1/4}\sqrt{H_{\rm dec}M_G}$) can be
less than 1 MeV, which is forbidden by the BBN constraint if $Q$-balls
dominate the Universe before their decay.  Thus, we have another
constraint on the reheating temperature as
\begin{equation}
T_R \gtrsim 0.4 \times 10^{-6}A_T^{1/4}A_{\rm dec}^{1/8}\left(\frac{|K|({\bar \epsilon}\beta)}{\alpha_g^4}\right)^{1/4}\left(\frac{\phi_{\rm osc}}{M_G}\right)^{3/2}
\left(\frac{10^3 {\rm GeV}}{m_\phi}\right)^{1/4}M_G \equiv T_R^{c,BBN} \label{trcbbngr}
\end{equation}
for Cases 1 and 3. 

Using Eqs.~\eqref{th_omgwf}, \eqref{th_f}, \eqref{parawithtr},
\eqref{hdomposk}, \eqref{dilute}, \eqref{redshift} with the decay time
Eq.~\eqref{dechnegk}, the present density parameter of the GWs,
$\Omega_{\rm GW}^0$, is given by
\begin{align}
\Omega_{\rm GW}^0&\simeq\left\{
\begin{array}{ll}
\dfrac{1}{216}\left(\dfrac{3}{\pi}\right)^{2/3}\dfrac{\eta^{4/3}\alpha_g^4}{\alpha^{1/3}A_{\rm dom}^{1/3} A_T^{2/3}|K|^{2/3}} ({\bar \epsilon}\beta)^{-1} 
\dfrac{M_G^{2/3}T_R^4}{m_\phi^{2/3}\phi_{\rm osc}^4} \dfrac{a_{eq}}{a_0} & \quad {\rm (Case \ 1, 3)},  \\ 
\dfrac{\alpha^{8/3}A_T^{2/3}}{54\alpha_g^{4/3}} ({\bar \epsilon}\beta) 
\left(\dfrac{\phi_{\rm osc}}{M_G}\right)^{16/3}\dfrac{a_{eq}}{a_0} & \quad {\rm (Case \ 2,4)},
\end{array} \right. \label{omgwnegk}
\end{align}
and the frequency, $f_0$, is given by 
\begin{align}
f_0&\simeq\left\{
\begin{array}{ll}
\dfrac{\sqrt{3}}{2\pi}\left(\dfrac{3}{\pi}\right)^{1/6}\dfrac{ \eta^{1/3} \alpha_g^2 
A_{\rm dom}^{1/4}A_0^{1/3}}{\alpha^{1/12} A_{\rm dec}^{1/12}A_T^{3/4}|K|^{1/6}}
({\bar \epsilon}\beta)^{-1/2} \dfrac{T_R^2
M_G^{7/6}}{m_\phi^{1/6}\phi_{\rm osc}^3} T_0 & \quad {\rm (Case \ 1, 3)}, \\ 
\dfrac{\sqrt{6}\alpha_g^{2/3}\alpha^{2/3}A_0^{1/3}}{2 \pi A_T^{1/6}} 
\dfrac{T_R}{\phi_{\rm osc}^{2/3}M_G^{1/3}}T_0 & \quad {\rm (Case \ 2,4)}.  
\end{array} \right.\label{f0negk}
\end{align}
From Eqs.~\eqref{tcposk}, \eqref{hdomposk}, and \eqref{dechnegk}, the critical temperatures $T_R^{c1}$ and $T_R^{c2}$ in this
case are given by
\begin{align}
T_R^{c1} & =\dfrac{A_T^{1/8}}{\alpha_g^{3/4}}({\bar \epsilon}\beta)^{1/8}\dfrac{m_\phi^{1/2}\phi_{\rm osc}^{3/4}}{M_G^{1/4}}, \\
T_R^{c2} & = \left(\dfrac{8\pi |K|}{3}\right)^{1/6} \dfrac{\alpha^{3/4}A_{\rm dom}^{1/12}A_T^{1/3}}{\eta^{1/3}\alpha_g^{4/3}} ({\bar \epsilon}\beta)^{1/2} \dfrac{m_\phi^{1/6}\phi_{\rm osc}^{7/3}}{M_G^{3/2}},  
\end{align}
which characterize all cases as summarized in Table \ref{Table:2}.

\begin{figure}[htbp]
 \begin{center}
  \includegraphics[width=150mm]{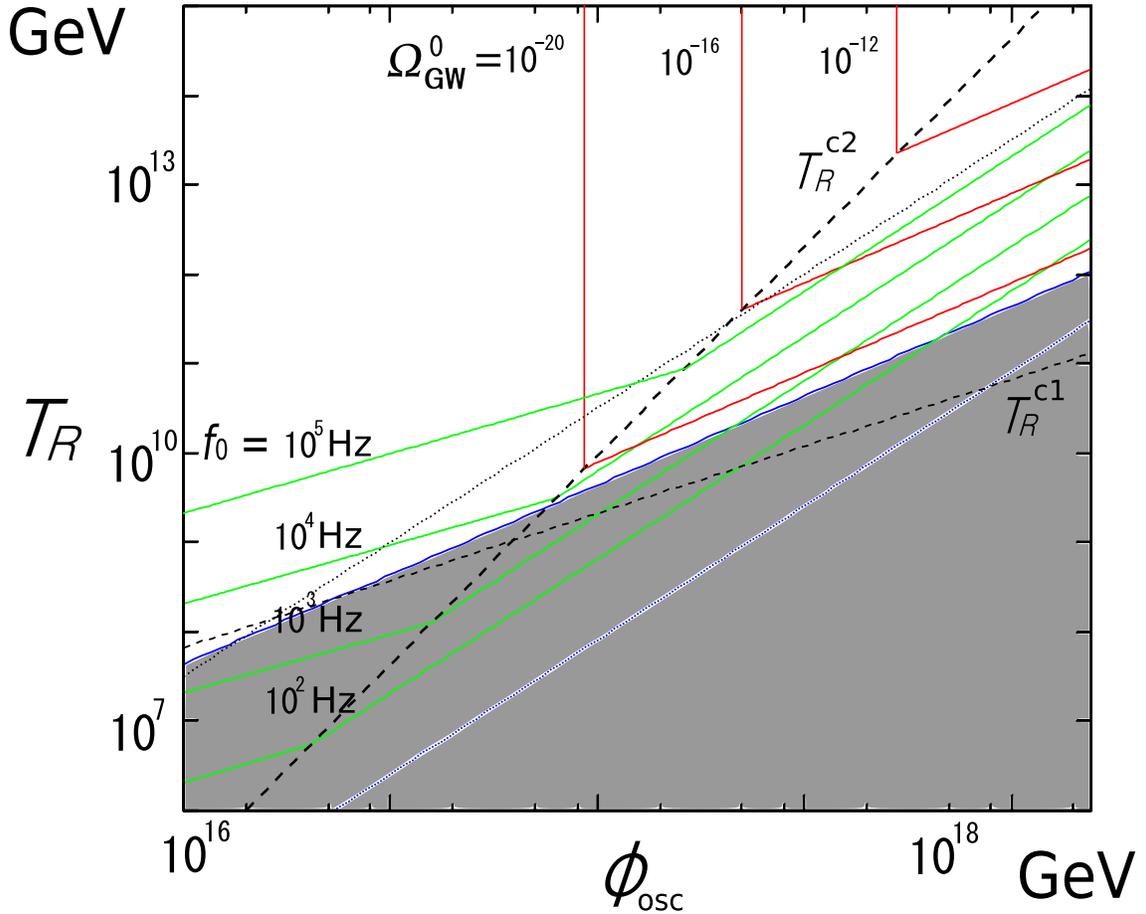}
 \end{center}
 \caption{The same plot as Fig.\ref{fig:contour1} but for the case of the gravity-mediated 
 SUSY-breaking model with negative $K$ is shown. The shaded region represents the forbidden region where 
$T_R<T_R^c$ or $T_R<T_R^{c,BBN}$.  The parameters are taken to be  
$\eta\simeq 1, \alpha \simeq 1$, $\alpha_g\simeq |K| \simeq 0.1$ and 
$m_{\phi}\simeq 10^2 {\rm GeV}$, and 
the effective relativistic degrees of freedom of MSSM are assumed for $A_T, A_{\rm dom}, A_{\rm dec}$ 
and $A_0$.}
 \label{fig:contour2}
\end{figure}

Figure \ref{fig:contour2} shows the contour plot of the present density
parameter of the GWs, $\Omega_{\rm GW}^0$, and their frequency, $f_0$.
$\Omega_{\rm GW}^0$ is independent of the reheating temperature, $T_R$,
in Cases 2 and 4 while it depends on both $T_R$ and $\phi_{\rm osc}$ in
Cases 1 and 3.
$f_0$ is sensitive to $T_R$. 
We find that it is almost impossible to detect the GWs from $Q$-ball
formation even by the future detectors in this case, as with the case
with positive $K$.  The relatively small decay rate of the $Q$-balls leads
to a lower frequency of the GWs, but it also results in the $Q$-ball
domination, which dilutes GWs.

We comment on the possibility of the $Q$-ball evaporation.  In this case
we need to consider the $Q$-ball evaporation both in the thermal log case
[Eq. \eqref{evthq1}] and in the gravity-mediation case
[Eq. \eqref{evgrq1}].  The condition for the evaporation of the $Q$-balls
before their decay in the case when the thermal logarithmic term
dominates the potential is the same as that for positive $K$.
Using Eq.~\eqref{thq}, the condition for the $Q$-ball evaporation after the $Q$-ball transformation is 
\begin{equation}
\phi_{\rm osc}<3 \times \frac{T_R^{2/3}M_G^{13/27}}{m_\phi^{4/27}} \simeq 10^{15}\left(\frac{T_R}{10^9{\rm GeV}}\right)^{2/3} \left(\frac{m_\phi}{1{\rm TeV}}\right)^{-4/27} \equiv \phi_{\rm osc}^{\rm evap}. 
\end{equation}
As in the case with positive $K$, we find from Fig. \ref{fig:contour2}  
that the evaporation of $Q$-balls can modify the above estimation of $\Omega_{GW}^0$ and $f_0$ 
only when  $f_0$ is too large ($f_0>10^5$ Hz). 
Therefore, even if we take the evaporation effects into account, the
conclusion is unchanged.

Summarizing our results in the gravity or anomaly mediated SUSY-breaking
models, we find that it is possible to generate a large amount of GWs
{}from the $Q$-ball formation, but the present frequency is relatively high. This is
mainly because the initial frequencies of such GWs are rather large and
cannot be redshifted enough.  Thus, we conclude that in this mediation
mechanism, the GWs from the $Q$-ball formation cannot be detected by the
future detectors (designed or planned) even if thermal log terms induce
the $Q$-ball formation.

\subsection{Gauge-mediated SUSY-breaking model}

\label{gaugecase} 

In this subsection, we study the present properties of the GWs from the
$Q$-ball formation in the case where the $Q$-balls are formed when the
thermal logarithmic term is the dominant contribution to the potential
of the AD field in the gauge-mediated SUSY-breaking model.

{}From the condition that the thermal logarithmic term dominates the
potential of the AD field, Eqs.~\eqref{th_ow_gauge}, we have the lower
bound on the reheating temperature
\begin{equation}
T_R>{\rm max}\left\{\frac{A_T^{1/4}}{\alpha_g}\left(\frac{\phi_{\rm osc}}{M_G}\right)^{1/2}M_F, A_T^{1/4}\left(\frac{m_{3/2}}{\alpha_g^2 M_G}\right)^{1/2}\phi_{\rm osc}\right\}, \label{gentrcongauge}
\end{equation}  
where we have used the relation Eq. \eqref{parawithtr}.  This bound
strongly constrains the frequencies of the produced GWs.  As mentioned
before, there are three possibilities of the $Q$-ball transformation,
Cases A, B, and C (Fig. \ref{fig:gauge}), depending on the dominant term
($V_{\rm gauge}$ or $V_{\rm grav2}$) in the potential and on the sign of
$K$.  We estimate the present properties of the GWs from the $Q$-ball
formation in each case.

\subsubsection{Case A: $V_{\rm gauge}$ driven $Q$-ball transformation}
\label{subsec:thgauge}

First we consider Case A, in which the $Q$-balls are formed by the thermal
logarithmic potential and their type is changed into the gauge-mediated
type at the critical temperature $T_c^A$.

From Eqs.~\eqref{decga} and \eqref{thq}, the Hubble parameter at the $Q$-ball decay is given by
\begin{align}
H_{\rm dec}&=\dfrac{\sqrt{2}\pi^2}{48}({\bar \epsilon}\beta)^{-5/4}\left(\dfrac{\alpha_g^4 }{A_T}\right)^{5/4}\dfrac{M_G^{5/2}T_R^5M_F}{\phi_{\rm osc}^{15/2}}. \label{hdecga} 
\end{align}
We have three constraints on the parameters of this model.  The
first constraint comes from the BBN.  For the successful BBN, the
$Q$-balls must decay before BBN and the decay temperature should be larger than $1 {\rm MeV}$ 
if $Q$-balls dominate the energy density of the Universe. This requirement becomes
\begin{equation}
T_R \gtrsim 3 \times 10^{-9} ({\bar \epsilon}\beta)^{1/4}\frac{A_T^{1/4}A_{\rm dec}^{1/10}}{\alpha_g}\frac{\phi_{\rm osc}^{3/2}}{M_G^{3/10}M_F^{1/5}} \equiv T_R^{c,BBN}. \label{contrgaubbn}
\end{equation}
The second one is the condition for $V_{\rm thermal}$ to dominate the
potential of the AD field Eq. \eqref{gentrcongauge},
\begin{equation}
T_R>\frac{A_T^{1/4}}{\alpha_g}\left(\frac{\phi_{\rm osc}}{M_G}\right)^{1/2}M_F \equiv T_R^{c,{\rm th}}. 
\end{equation}
The last one is the condition that $V_{\rm gauge}>V_{\rm grav2}$ at $\phi=\phi_c$, 
\begin{equation}
T_R>\frac{A_T^{1/4}}{\alpha_g}({\bar \epsilon}\beta)^{1/4}\frac{m_{3/2}\phi_{\rm osc}^{3/2}}{M_G^{1/2}M_F} \equiv T_R^{c,{\rm gr}}.  \label{trvavr} 
\end{equation}  
Here we have used Eqs.~\eqref{thphiq} and \eqref{tca}. 

From Eqs.~\eqref{t*}, \eqref{omegaq*}, \eqref{tca}, \eqref{parawithtr}, and \eqref{hdomgene}, we obtain the Hubble parameter at the $Q$-ball domination as
\begin{equation}
H_{\rm dom}=\frac{\alpha^{9/2}A_{\rm dom}^{1/2}A_T^{1/2}}{9\eta^2\alpha_g^2}({\bar \epsilon}\beta)^{3/2} 
\dfrac{\phi_{\rm osc}^5 M_F^2}{M_G^6}. \label{hdomga}
\end{equation}
The present amount of the GWs from the $Q$-ball formation is then given by 
\begin{align}
\Omega_{\rm GW}^0&\simeq\left\{
\begin{array}{ll}
\dfrac{1}{216}\left(\dfrac{3 \pi^2}{\sqrt{2}}\right)^{2/3}\dfrac{\eta^{4/3}\alpha_g^{10/3}}{\alpha^{1/3} 
A_{\rm dom}^{1/3}A_T^{1/2}}({\bar \epsilon}\beta)^{-5/6}\dfrac{T_R^{10/3} M_G^{1/3}}{\phi_{\rm osc}^3 M_F^{2/3}}\dfrac{a_{eq}}{a_0} & \quad {\rm (Case \ 1,3)}, \\ 
\dfrac{\alpha^{8/3}A_T^{2/3}}{54\alpha_g^{4/3}}({\bar \epsilon}\beta) \left(\dfrac{\phi_{\rm osc}}{M_G}\right)^{16/3}\dfrac{a_{eq}}{a_0} & \quad {\rm (Case \ 2,4)},
\end{array} \right. \label{omgwgauge3}
\end{align}
and the frequency, $f_0$, is estimated as
\begin{align}
f_0&\simeq\left\{ 
\begin{array}{ll}
 \dfrac{\sqrt{3}}{2\pi} \left(\dfrac{3\pi^2}{\sqrt{2}}\right)^{1/6} \dfrac{\eta^{1/3}\alpha_g^{11/6}  A_0^{1/3}}{\alpha^{1/12} 
A_{\rm dec}^{1/12}A_T^{11/24}}({\bar \epsilon}\beta)^{-11/24}
\dfrac{M_G^{13/12}T_R^{11/6}}{\phi_{\rm osc}^{11/4}M_F^{1/6}}T_0 & \quad {\rm (Case \ 1, 3)}, \\ 
\dfrac{\sqrt{6}\alpha_g^{2/3}\alpha^{2/3}A_0^{1/3}}{2 \pi A_T^{1/6}} \dfrac{T_R}{\phi_{\rm osc}^{2/3}M_G^{1/3}}T_0 & \quad {\rm (Case \ 2,4)}.  
\end{array} \right.\label{f0gauge3}
\end{align} 
Here we have used Eqs.~\eqref{th_omgwf}, \eqref{th_f},  \eqref{parawithtr}, \eqref{dilute}, \eqref{redshift}, \eqref{hdecga}, and \eqref{hdomga}.   
The critical temperatures, $T_R^{c1}$ and $T_R^{c2}$, using Eqs.~\eqref{tca}, \eqref{hdecga}, and \eqref{hdomga}, can be estimated as  
\begin{align}
T_R^{c1} & = \dfrac{M_F}{\alpha_g^{1/2}},  \label{trgagr1}\\
T_R^{c2} & = \left(\frac{8\sqrt{2}}{3\pi^2}\right)^{1/5}({\bar \epsilon}\beta)^{11/20}\frac{\alpha^{9/10}A_T^{7/20}A_{\rm dom}^{1/10}}{\alpha_g^{7/5}\eta^{2/5}}\frac{M_F^{1/5} \phi_{\rm osc}^{5/2}}{M_G^{17/10}}, \label{trgagr2}
\end{align}
which characterize all four cases as summarized in Table \ref{Table:2}. 

The solid (red) line in Fig. \ref{fig:fr_om} represents the maximal amplitudes of the
GWs, $\Omega_{\rm GW}^{0, {\rm max}}$, from $Q$-ball formation at the
frequency ranging from $1$ Hz to $10^9$ Hz in Case A.  When $f_0
\lesssim 10^{2}$ Hz, $\Omega_{\rm GW}^{0,{\rm max}} \propto f_0^{5/2}$, 
which corresponds to the parameter region with $T_R=T_R^{c,{\rm
BBN}}$, $10^4 {\rm GeV}<M_F<10^{10}{\rm GeV}$, and $\phi_{\rm osc}=M_G$.  
When $10^{2}{\rm Hz} \lesssim f_0 \lesssim 10^{8}$ Hz, 
$\Omega_{\rm GW}^{0,{\rm max}} \propto
f_0^{20/11}$, which corresponds to the parameter region with
$T_R^{c,{\rm BBN}}<T_R<T_R^{c2}$, $\phi_{\rm osc}=M_G$, and $M_F=10^4$ GeV. Note
that in both cases, there is the $Q$-ball dominated era.  When $f_0
\gtrsim 10^{8}$ Hz, $\Omega_{\rm GW}^{0,{\rm max}}$ is obtained for
$\phi_{\rm osc}=M_G$ without the $Q$-ball dominated era.  As is seen in
Fig. \ref{fig:fr_om}, $\Omega_{\rm GW}^{0,{\rm max}}$ becomes larger
than $10^{-16}$ but at $f_0 \gtrsim 10^3$ Hz, which goes outside the
sensitivity range of the DECIGO or BBO.  Therefore, it is almost
impossible to detect the GWs from $Q$-ball formation even by the future
detectors in this case.

We comment on the possibility of the $Q$-ball evaporation.  
In this case, $Q$-balls can evaporate out before the decay evaluated above 
and the estimation of $\Omega_{GW}^0$ and $f_0$ can break down. 
However, this is the case only for $f_0>10^6$ Hz. 
Thus, the above conclusion is unchanged by the evaporation effects.

\subsubsection{Case B: $V_{\rm grav} (K < 0) $ driven $Q$-ball transformation}

Next we consider Case B. In this case, $Q$-balls are formed by the thermal
logarithmic potential and change their type into the gravity-mediated
type ($K < 0$) at the critical temperature $T_c^B$.  This case is
similar to the case of the gravity-mediated SUSY-breaking model with
negative $K$.  Thus, we need only to replace $m_\phi$ by
$m_{3/2}(<m_\phi)$.  While the larger amount of the GWs is
obtained by replacing $m_\phi$ by $m_{3/2}(<m_\phi)$ of Case 3 in
Eq. \eqref{f0negk}, which is favorable for the detection, the present
frequency of the GWs is also enhanced, which is unfavorable for the
detection. 

There are also three constraints on the reheating temperature. 
The first one is the BBN constraint, which is required  when $Q$-balls
dominate the energy density of the Universe, and is given by
\begin{equation}
 T_R \gtrsim 2\times 10^{-6}A_T^{1/4}A_{\rm dec}^{1/8}\left(\frac{|K|({\bar \epsilon}\beta)}{\alpha_g^4}\right)^{1/4}\left(\frac{\phi_{\rm osc}}{M_G}\right)^{3/2}\left(\frac{1 {\rm GeV}}{m_{3/2}}\right)^{1/4}M_G \equiv T_R^{c,BBN}. \label{tctc}  
\end{equation}
This constraint can be derived in the same way as deriving Eq.~\eqref{trcbbngr}. 
The second one is the condition that the thermal logarithmic potential should dominate the 
potential of the AD field at the $Q$-ball formation  and is given by [Eq. \eqref{gentrcongauge}], 
\begin{equation}
T_R>A_T^{1/4}\left(\frac{m_{3/2}}{\alpha_g^2 M_G}\right)^{1/2}\phi_{\rm osc}\equiv T_R^{c,{\rm th}}.     \label{th_ow_grav3}  
\end{equation}
The last one is the condition for $V_{\rm grav}$ to dominate $V_{\rm gauge}$ at the $Q$-ball transformation, 
\begin{equation}
T_R<A_T^{1/4}\alpha_g^{-1} ({\bar \epsilon}\beta)^{1/4}\frac{m_{3/2}\phi_{\rm osc}^{3/2}}{M_FM_G^{1/2}}.  \label{trcgr}
\end{equation}
Here we have used Eqs.~\eqref{thphiq} and \eqref{critemnew}. 
Since $m_{3/2}/M_F\simeq M_F/M_G \lesssim 10^{-8}$ if there is no large
hierarchy between the SUSY-breaking sector and the messenger sector, the
first constraint contradicts with the last one.
However, if there is a hierarchy between them, three conditions can be made compatible. 

The dashed (green) line in Fig. \ref{fig:fr_om} represents the maximal amount  
of the GWs, $\Omega_{\rm GW}^{0, {\rm max}}$, from the $Q$-ball formation at the frequency
ranging from $1$ Hz to $10^9$ Hz in Case B.  When $f_0
\lesssim 10^{4}$ Hz, $\Omega_{\rm GW}^{0,{\rm max}} \propto f_0^{5/2}$,
which corresponds to the parameter region with $T_R=T_R^{c,{\rm BBN}}$,
$m_{3/2}<10$ GeV, and $\phi_{\rm osc}=M_G$. For $10^{4}{\rm Hz}
\lesssim f_0 \lesssim 10^{9}$ Hz, $\Omega_{\rm GW}^{0,{\rm max}} \propto
f_0^{20/11}$, which corresponds to the parameter region with
$T_R=T_R^{c,{\rm gr}}$, $\phi_{\rm osc}=M_G$, and $m_{3/2}<10$ GeV.
Note that in both cases, there is the $Q$-ball dominated era.  For $f_0
\gtrsim 10^{9}$ Hz, $\Omega_{\rm GW}^{0,{\rm max}}$ is obtained for
$\phi_{\rm osc}=M_G$ without the $Q$-ball dominated era.  As  seen in
Fig. \ref{fig:fr_om}, $\Omega_{\rm GW}^{0,{\rm max}}$ is larger than
$10^{-16}$ for $f_0\simg 10^3$ Hz, which again goes outside the
sensitivity range of the DECIGO or BBO.  Therefore, it is almost
impossible to detect the GWs from $Q$-ball formation even by the future
detectors in this case too.  The conclusion is unchanged by the
evaporation effects for the same reason discussed in Sec. \ref{subsec:thgauge}.
In this case, the $Q$-ball evaporation out takes place when $f_0>10^{3\sim 4}$ Hz.

\subsubsection{Case C: $V_{\rm grav2} (K>0) $ driven $Q$-ball transformation}
\label{subsubsec:C}

Finally, we consider Case C.  In this case, $Q$-balls are formed by the
thermal logarithmic potential and are transformed into the almost
homogeneous oscillating AD field at the critical temperature $T_c^B$.
Afterwards, the potential energy of the AD field decreases and $V_{\rm
gauge}$ dominates the potential finally. At that time, $Q$-balls are
formed again.  Note that the second $Q$-balls are of the ``delayed'' type
and hence the GWs from the second $Q$-ball formation cannot be detected,
as we will see in Appendix. B.

From Eq.~\eqref{decga}, the Hubble parameter at the $Q$-ball decay
is given by 
\beqa 
H_{\rm dec}=\dfrac{\pi^2}{24\sqrt{2}}\dfrac{m_{3/2}^5}{M_F^4}. 
\label{hdecgac} 
\eeqa
We have again three constraints on the parameters of this model. One
comes from the BBN constraint,
\begin{equation}
m_{3/2}>4 \times 10^{-9}A_{\rm dec}^{1/20}M_F^{4/5}M_G^{1/5}\equiv m_{3/2}^{\rm min},  \label{contrgaubbn1}
\end{equation}
which is required when the $Q$-balls dominate the energy density of the
Universe.  Another is the condition for $V_{\rm thermal}$ to dominate
the potential of the AD field, which coincides with
Eq.~\eqref{th_ow_grav3} ,
\begin{equation}
T_R>A_T^{1/4}\left(\frac{m_{3/2}}{\alpha_g^2 M_G}\right)^{1/2}\phi_{\rm osc}\equiv T_R^{c,{\rm th}}. \label{trvv}
\end{equation}
The last one is the condition that $V_{\rm grav}>V_{\rm gauge}$ at
$\phi=\phi_c$, which coincides with Eq.~\eqref{trcgr}, 
\begin{equation}
T_R<\frac{A_T^{1/4}}{\alpha_g }({\bar \epsilon}\beta)^{1/4}\frac{m_{3/2}\phi_{\rm osc}^{3/2}}{M_G^{1/2}M_F} \equiv T_R^{c,{\rm gr}}.  \label{trvv1} 
\end{equation} 
{}From Eqs. \eqref{trvv} and \eqref{trvv1}, we have the inequality 
\begin{equation}
\phi_{\rm osc}>({\bar \epsilon}\beta)^{-1}\frac{M_F^2}{m_{3/2}}. \label{ineqthposk}
\end{equation}
If there is no hierarchy between the SUSY-breaking sector and the messenger sector, 
$M_F^2\simeq m_{3/2}M_G$, the inequality \eqref{ineqthposk} 
cannot be satisfied. Hence we consider the case when there is some hierarchy between them 
so that Eq. \eqref{ineqthposk} is satisfied. 
 
The Hubble parameter at the $Q$-ball transformation is given by 
\begin{equation}
H_{\rm dom}\simeq \frac{\alpha^{9/2}A_{\rm dom}^{1/2}A_T}{9\eta^2 \alpha_g^4}({\bar \epsilon}\beta)^{2} 
\dfrac{m_{3/2}^2 \phi_{\rm osc}^8}{M_G^7T_R^2}, \label{hdomgac}
\end{equation}
which is obtained by replacing $m_\phi$ in Eq.~\eqref{hdomposk} by $m_{3/2}$. 
The present amount of the GWs from the $Q$-ball formation is then given by 
\begin{align}
\Omega_{\rm GW}^0&\simeq\left\{
\begin{array}{ll}
\dfrac{1}{216}\left(\dfrac{3 \pi^2}{\sqrt{2}}\right)^{2/3}\dfrac{\eta^{4/3} 
\alpha_g^{4/3}}{\alpha^{1/3} A_{\rm dom}^{1/3}}({\bar \epsilon}\beta)^{-1/3}\dfrac{T_R^{4/3}m_{3/2}^{2}}{M_G^{2/3} M_F^{8/3}}\dfrac{a_{eq}}{a_0} & \quad {\rm (Case \ 1,3)}, \\ 
\dfrac{\alpha^{8/3}A_T^{2/3}}{54\alpha_g^{4/3}}({\bar \epsilon}\beta) \left(\dfrac{\phi_{\rm osc}}{M_G}\right)^{16/3}\dfrac{a_{eq}}{a_0} & \quad {\rm (Case \ 2,4)},
\end{array} \right. \label{omgwnegk3}
\end{align}
and the frequency, $f_0$, is estimated as  
\begin{align}
f_0&\simeq\left\{
\begin{array}{ll}
 \dfrac{\sqrt{3}}{2\pi} \left(\dfrac{3\pi^2}{\sqrt{2}}\right)^{1/6} \dfrac{\alpha_g^{4/3}  A_0^{1/3}\eta^{1/3}}
{\alpha^{1/12} A_{\rm dec}^{1/12}A_T^{1/3}({\bar \epsilon}\beta)^{1/3}}\dfrac{T_R^{4/3}m_{3/2}^{1/2}M_G^{5/6}}{\phi_{\rm osc}^{2}M_F^{2/3}}T_0 & \quad {\rm (Case \ 1,3)}, \\ 
\dfrac{\sqrt{6}\alpha_g^{2/3}\alpha^{2/3}A_0^{1/3}}{2 \pi A_T^{1/6}} \dfrac{T_R}{\phi_{\rm osc}^{2/3}M_G^{1/3}}T_0 & \quad {\rm (Case \ 2,4)}.  
\end{array} \right.\label{f0negk3}
\end{align} 
Here we have used Eqs.~\eqref{th_omgwf}, \eqref{th_f},
\eqref{parawithtr}, \eqref{dilute}, \eqref{redshift}, \eqref{hdecgac},
and \eqref{hdomgac}.  From Eqs.~\eqref{critemnew}, \eqref{hdecgac}, and
\eqref{hdomgac}, the critical temperatures, $T_R^{c1}$ and $T_R^{c2}$,
are estimated as
\begin{align}
T_R^{c1} & = \left(\frac{A_T({\bar \epsilon}\beta)}{\alpha_g^6}\right)^{1/8}\frac{m_{3/2}^{1/2}\phi_{\rm osc}^{3/4}}{M_G^{1/4}},  \label{trcc1}\\
T_R^{c2} & = \left(\frac{8\sqrt{2}}{3\pi^2}\right)^{1/2}({\bar \epsilon}\beta)\frac{\alpha^{9/4}A_{\rm dom}^{1/4}A_T^{1/2}}{\alpha_g^2 \eta}\frac{M_F^2 \phi_{\rm osc}^{4}}{M_G^{7/2} m_{3/2}^{3/2}}, \label{trcc2}
\end{align}
which characterize all four cases as summarized in Table \ref{Table:2}. 

\begin{figure}[htbp]
\begin{center}
  \includegraphics[width=150mm]{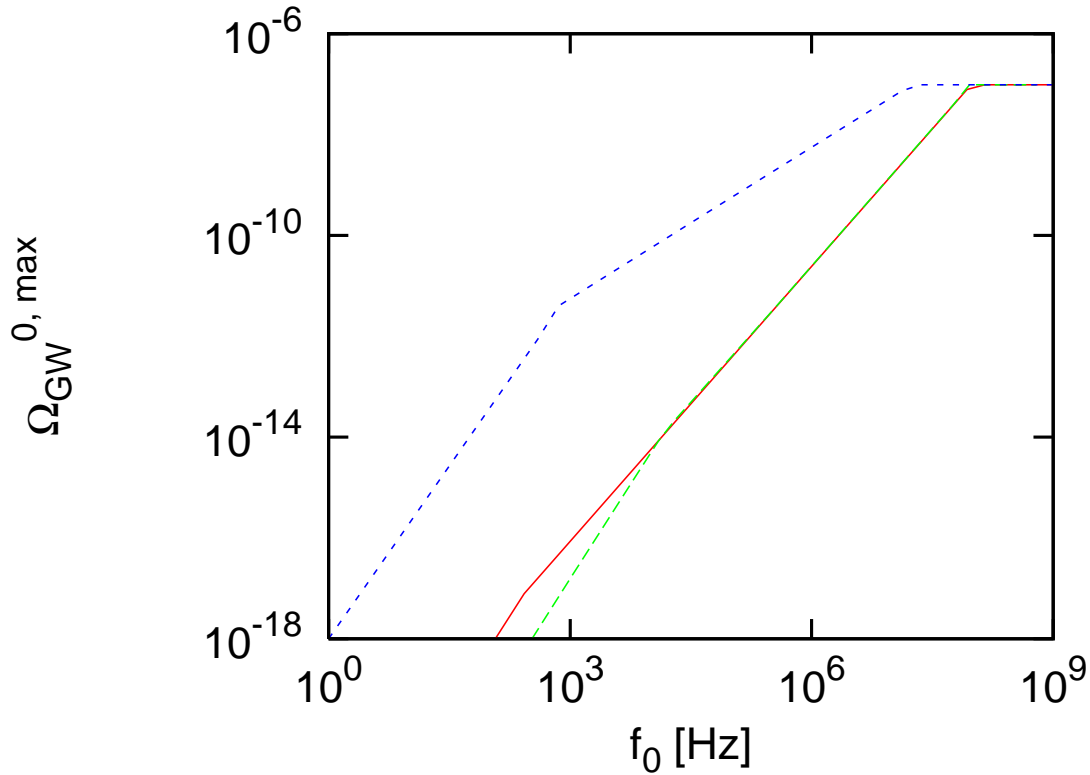} 
\caption{The maximal amplitude of the GWs from $Q$-ball formation at the
frequency ranging from 1 Hz to $10^9$ Hz in each
 case is shown [Case A (red solid line), Case B (green dashed line) and
 Case C (blue dotted line)].  The horizontal axis represents the
 frequency, $f_0$, and the vertical axis represents the present density
 parameter of the GWs, $\Omega_{\rm GW}^{0,{\rm max}}$. The parameters are taken to be 
$\eta \simeq 1, \alpha \simeq 1$, and $h\simeq \alpha_g\simeq |K| \simeq 0.1$, and 
 the effective relativistic degrees of freedom  of MSSM are assumed.} \label{fig:fr_om}
  \end{center}
\end{figure}

The dotted (blue) line in Fig. \ref{fig:fr_om} is the maximal
amplitudes of the GWs, $\Omega_{\rm GW}^{0, {\rm max}}$, from the $Q$-ball
formation at the frequency ranging from 1 Hz to $10^9$ Hz in Case C.
For $f_0\lesssim 10^3$ Hz, $\Omega_{\rm GW}^{0,{\rm max}} \propto
f_0^{16/7}$, which corresponds to the parameter region with $M_F\simeq
10^4$ GeV, $m_{3/2}<10$ GeV and $T_R=T_R^{c,{\rm th}}$ for $\phi_{\rm
osc}=M_G$.  For $10^3$ Hz $\lesssim f_0\lesssim 10^{7}$ Hz, $\Omega_{\rm
GW}^{0,{\rm max}} \propto f_0$, which corresponds to the parameter
region with $M_F\simeq 10^4$ GeV, $\phi_{\rm osc} =M_G$ and
$T_R>T_R^{c,{\rm th}}$ for $m_{3/2}=10$ GeV.  Note that in both cases,
there is the $Q$-ball dominated era.  For $f_0> 10^7$ Hz, $\Omega_{\rm
GW}^{0,{\rm max}}$ is obtained for $\phi_{\rm osc}=M_G$ and there is no
$Q$-ball dominated era.  As seen in Fig. \ref{fig:fr_om}, $\Omega_{\rm
GW}^{0,{\rm max}}$ is larger than $10^{-16}$ for $f_0\simg 10$ Hz, which
is on the edge of the DECIGO or BBO sensitivity range.  Thus, it is
difficult but not impossible to detect such GWs by the next generation
detectors.  The parameters that realizes $\Omega_{\rm GW}^0\simeq
10^{-16}$ at $f_0\simeq 10$ Hz are given by
\begin{equation}
M_F\simeq10^4 {\rm GeV}, \ \ m_{3/2}\simeq 10 {\rm GeV}, \ \ \phi_{\rm osc} \simeq M_G \ \ {\rm and} \ \ T_R =  T_R^{c,{\rm th}}\simeq 10^{10}{\rm GeV}. 
\end{equation}
The conclusion above is unchanged by the evaporation effects for the
same reason discussed in Sec. \ref{subsec:thgauge}.  In this case, the
$Q$-ball evaporation takes place when $f_0>10$ Hz.  One should notice
that though gravitino does not overclose the Universe because of large
entropy production from $Q$-ball decay, the next-to-lightest
supersymmetric particle (NLSP) decay may spoil the success of the BBN in
some cases since the hadronic decay product of NLSP would destroy the
light elements \cite{gravitinoprb}.

Here we comment on the present baryon and lepton asymmetry.  In the
situation where the GWs from the $Q$-ball formation might be detected,
both present baryon/lepton asymmetry and radiation are generated by the
$Q$-ball decay. The produced baryon/lepton asymmetry from the $Q$-ball decay
would be rather large and be of the order of unity unless the parameter
in the $A$-term, $a_m$, is strongly suppressed \cite{Kawasaki:2002hq}.
In the case of $Q$-ball with baryonic charge, it is far beyond the
experimental bound on the present baryon asymmetry.  Even in the
case of $L$-balls, that is, $Q$-balls with lepton charges but without baryon
charge, the situation does not change since the decay temperature of
$Q$-ball is about 500 GeV and hence the lepton asymmetry is converted to
baryon asymmetry by sphaleron process \cite{Kuzmin:1985mm} that
conserves $B-L$ charge.  The way to avoid such large baryon/lepton
asymmetry in this scenario is that the $Q$-balls are made of the AD field
with $B-L=0$. 
In this case, however, we need other baryogenesis mechanisms.

\section{Discussion and Conclusions}

\label{sec:DC}

In this paper, we have discussed the detectability of the GWs from the
$Q$-ball formation.  At the $Q$-ball formation, $Q$-balls with large $Q$ can
produce a large amount of GWs.
However, such $Q$-balls decay slowly and they may dominate the energy
density of the Universe so that GWs are significantly diluted.
Therefore the detectability of the GWs is determined by these two
competing effects.

We have concentrated on the case of thermal log type $Q$-balls. Such
$Q$-balls do not dominate the energy density of the Universe until the
dominant potential for the AD field changes, after that their properties
are changed.  We have shown that in the gauge-mediated SUSY-breaking
model, if the reheating temperature is $T_R\simeq 10^{10}$ GeV and the
initial field value of the AD field is $\phi_{\rm osc} \simeq M_G$ with
$m_{3/2}\simeq 10 {\rm GeV}$ and $M_F \simeq 10^4$ GeV, the present
density parameter of the GWs from the $Q$-ball formation can be as large
as $\Omega_{\rm GW}^0 \simeq 10^{-16}$ and their frequency is $f_0
\simeq 10$ Hz.  Thus, it is difficult but not impossible to detect them
by next-generation gravitational detectors like DECIGO or BBO, but the
parameter region for detectable GWs is very small.  Moreover, we have
shown that it is almost impossible to detect GWs from $Q$-ball formation
in other cases. In other cases when the thermal logarithmic potential
drives the $Q$-ball formation, though the present amounts of the GWs from
the $Q$-ball formation can be as large as $\Omega_{\rm GW}^0\simeq
10^{-8}$, the frequencies of such GWs are turned out to be very high.
In the cases where zero-temperature potential terms drive the $Q$-ball
formation, the present amount of the GWs from the $Q$-ball formation is
very small due to the large dilution.  Thus, the identification of such
GWs may determine the decay rate of inflaton or the initial condition of
the AD mechanism.  In Table \ref{Table:3}, we show the minimal frequency
of the GWs that satisfy $\Omega_{\rm GW}^0>10^{-16}$ in each thermal
dominated case, and the maximal amplitudes of GWs in each 
zero-temperature case are given in Table \ref{Table:4}.

\begin{table}[htbp]
\begin{center}
\begin{tabular}{l|c}\hline
Dominant term in the potential  & The smallest frequency $f_0$ that satisfies $\Omega_{\rm GW}^0>10^{-16}$   \\ \hline\hline
$V_{\rm thermal} \Rightarrow V_{\rm grav}(K>0)$ &  $10^{3-4}$ Hz\\ \hline
$V_{\rm thermal} \Rightarrow V_{\rm grav}(K<0)$ &  $10^{2-3}$ Hz\\ \hline
$V_{\rm thermal} \Rightarrow V_{\rm grav}(K>0)\Rightarrow V_{\rm gauge} $ & $10^{1-2}$ Hz \\ \hline
$V_{\rm thermal} \Rightarrow V_{\rm gauge} $ &   $10^{2-3}$ Hz \\ \hline
\end{tabular} 
\caption{The minimal frequency with $\Omega_{\rm GW}^0>10^{-16}$ for thermal log type $Q$-balls.\label{Table:3}}
\end{center}
\end{table}
\vspace{3mm}

\vspace{3mm}
\begin{table}[htbp]
\begin{center}
\begin{tabular}{l|c}\hline
Dominant term in the potential  & The maximal density parameter $\Omega_{\rm GW}^0$   \\ \hline\hline
$V_{\rm grav}(K<0) $ & $10^{-25}$ \\ \hline
$V_{\rm gauge} $ & $10^{-21}$\\ \hline
$V_{\rm grav}(K>0)\Rightarrow V_{\rm gauge} $ &  $10^{-24}$\\ \hline
$V_{\rm grav}(K>0)\Rightarrow V_{\rm thermal} \Rightarrow V_{\rm gauge} $  & $10^{-24}$ \\  \hline
$V_{\rm thermal}(c<0) \Rightarrow V_{\rm gauge} $ & $10^{-24}$\\ \hline
\end{tabular}
\caption{The maximal density parameter for $Q$-balls with the zero-temperature potentials. \label{Table:4}}
\end{center}
\end{table}

We would like to comment on the difficulty in realizing the
successful parameter region. One difficulty is the NLSP decay that may spoil
the successful BBN when $m_{3/2}\simeq 10$ GeV, since the hadronic
energy release from NLSP would cause dissociation process of light
elements \cite{gravitinoprb}.  However, it can be avoided, for example,
if the mass of NLSP is heavy enough to decay quickly. Another is the
baryogenesis.  One may wonder whether the present baryon asymmetry can
be explained simultaneously for such parameter region for the detection
of the GWs.  Generally speaking, including the present case, the amount
of produced baryon asymmetry is typically large for the case that AD
condensates or $Q$-balls (almost) dominate the energy density of the
Universe so that the present radiations and baryons are attributed to
their decays. This is simply because the number densities of radiations
and baryons are of the same order unless the ($CP$-violating) $A$-terms are
suppressed by some symmetry.
Thus, once the GWs from the $Q$-ball formation are detected, we have the
following two possibilities. In the case that such $Q$-balls are
responsible for the present baryon asymmetry, the $A$-terms are suppressed
by symmetry reason. The second option is that $Q$-balls are irrelevant for
baryogenesis, which is realized for the AD fields with $B-L=0$.  

Although the detailed numerical calculations are required to compute the
spectrum of GWs associated with $Q$-ball formation, such GWs may be
differentiated from the GWs generated during inflation by their spectrum
and from the astrophysical origin like POP III stars by a
non-Gaussianity test \cite{Seto:2008xr}.  On the other hand, a first
order phase transition in the early Universe would produce similar
spectrum of GWs.  However, in our case, the gravitino mass must be
around $10$ GeV for the detection of the GWs from the $Q$-ball
formation.  As mentioned above, such a gravitino mass may induce
the NLSP decay problem.  Therefore, if collider experiments could
determine the gravitino mass by measuring the lifetime of the NLSP
\cite{stop_stau}, that would provide complemental information or even
rule out this scenario.

Although the identification is rather difficult, it is true that the
detection of such GWs by DECIGO or BBO gives us information of the early
Universe and the physics in the high energy scale, for example, it may
suggest that the AD mechanism with $B-L=0$ flat directions is
favored.\footnote{Baryogenesis using such flat directions is discussed
in the context of the spontaneous baryogenesis mechanism
\cite{spontaneous}.}

{\bf Note added in proof}: 
Recently Ref. \cite{Kasuya:2010vq} claims that $Q$ balls may survive even after $V_{\rm grav}$ with $K>0$ dominates
the potential.  It is true that a (thin-wall) $Q$-ball solution exists for such a potential, but it is still unclear whether such a configuration is realized in the expanding Universe. Furthermore, even if this is the case, the dilution factor may become larger so that the detectability of the GWs from $Q$-ball formation gets even worse. Thus, our conclusions are unchanged.

\acknowledgments 
We would like to thank Shinta Kasuya, Masahiro Kawasaki, 
Fuminobu Takahashi and Jun'ichi Yokoyama 
for useful comments.  This work was partially supported by JSPS (K.K.) 
and a Grant-in-Aid for Scientific Research 
{}from JSPS [No.\,20540280 (T.C.) and No.\,21740187 (M.Y.)] and from MEXT
[No.\,20040006 (T.C.)].  This work was also supported in part by Nihon
University and by Global COE Program (Global Center of Excellence for
Physical Sciences Frontier), MEXT, Japan.

\appendix

\section{charge evaporation from $Q$-balls}

As mentioned in Sec. \ref{subsec:Qfate}, $Q$-balls can release their
charges via the evaporation and the diffusion effects in the thermal
bath.  In this Appendix, we give the condition that $Q$-balls are
evaporated out by these effects before their decays.

At finite temperature, $Q$-balls are surrounded by charged free particles.
The minimum of free energy is achieved when all charges are distributed
in the form of free particles in the thermal plasma, which induces the
charge evaporation from the surfaces of $Q$-balls.

Charge emission of $Q$-ball consists of the two processes, evaporation at
low temperature and diffusion at high temperature.  When the difference
between the chemical potential of the plasma ($\mu_p$) and that of
$Q$-ball ($\mu_Q$) is small, chemical equilibrium is almost achieved and
hence the charge evaporation from the $Q$-ball surfaces is small.  Instead,
the charges in the plasmas around the $Q$-balls are taken away via the
diffusion process.  
The diffusion rate is given by
\begin{equation}
\Gamma_{\rm diff} \equiv \frac{dQ}{dt} \simeq -4\pi a_{\rm diff} T. 
\end{equation}
Here a numerical factor $a_{\rm diff}$ is estimated as $a_{\rm
diff}=4-6$ for quarks and squarks \cite{Joyce:1994zn,Davoudiasl:1998hg},
and $a_{\rm diff}=100-380$ for leptons and sleptons
\cite{Nelson:1991ab}.
 
On the other hand, when the difference between the chemical potential of
the plasma and that of $Q$-ball is large, $\mu_p\ll \mu_Q$, the
evaporation process from the $Q$-ball surface is active.  The evaporation
rate is estimated as
\begin{equation}
\Gamma_{\rm evap} \equiv \frac{dQ}{dt} \simeq -4\pi R_Q^2\xi(\mu_Q-\mu_p)T^2, 
\end{equation}
where
\begin{equation} 
\xi=\left\{
\begin{array}{ll}
1 & \quad \text{for} \ T>m_\phi, \\ 
\left(\dfrac{T}{m_\phi}\right)^2  & \quad \text{for} \ T < m_\phi.
\end{array} \right.
\end{equation}
Then, depending on the type of $Q$-ball, we obtain the evaporation rate,
\begin{equation}
\Gamma_{\rm evap} =\left\{
\begin{array}{ll}
-2\sqrt{2} \pi \xi Q^{1/4} \dfrac{T}{\alpha_g^{1/4}}  & {\rm for \ the \ thermal \ log \ type}, \\
- 8 \pi \xi \dfrac{T^2}{m_{\phi(3/2)}|K|} & {\rm for \ the \ gravity-mediated \ type}, \\
-2\sqrt{2} \pi \xi Q^{1/4} \dfrac{T^2}{M_F} & {\rm for \ the \ gauge-mediated \ type}. \\
\end{array}\right.
\end{equation}
Here we have used Eqs.~\eqref{qthermal}, \eqref{qgrav} and
\eqref{qthgauge} and the fact $\mu_Q\simeq \omega$.  The rate of
the charge transportation is determined by min$\{ |\Gamma_{\rm evap}|,
|\Gamma_{\rm diff}|\}$, because both the processes are necessary
to strip the charges from the $Q$-balls.  We define the transition
temperature as $T_{\rm tr}$, at which the diffusion process is more
effective than the evaporation process.  Hereafter we examine the
condition for each type that the $Q$-balls are evaporated out before their
decay.

\subsection{Thermal log type}

We consider the case where $V_{\rm thermal}$ dominates the potential.
Since the $Q$-ball transformation temperature satisfies $T_c>m_\phi$ in all
the cases and hence $\Gamma_{\rm evap}>\Gamma_{\rm diff}$ is always
satisfied, we have only to consider the diffusion process.
When the condition $H=\dfrac{1}{Q}|\Gamma_{\rm diff}|$ is satisfied at
$T>T_c$, $Q$-balls are evaporated out before their transformation, which is
realized if the following condition is satisfied: 
\begin{equation}
Q<\left\{
\begin{array}{ll}
\dfrac{4 \pi a_{\rm diff}}{A_T^{1/2}}\dfrac{T_R^2M_G}{T_c^3} \ \  &{\rm for}\ \   T_R<T_c ,\\
\dfrac{4\pi a_{\rm diff}}{A_T^{1/2}}\dfrac{M_G}{T_c} \ \  &{\rm for}  \ \ T_R>T_c . 
\end{array}\right. \label{evthq1}
\end{equation}
Here we have used the temperature dependence of the Hubble parameter:  
\begin{equation}
H\simeq \left\{
\begin{array}{ll}
\dfrac{A}{A_T^{1/2}}\dfrac{T^4}{T_R^2M_G} & {\rm for} \ \ T>T_R,  \\
A^{1/2}\dfrac{T^2}{M_G} & {\rm for} \ \ T_c<T<T_R. 
\end{array}\right.
\end{equation}

\subsection{Gravity-mediated type}

Next we consider the case where $V_{\rm grav(2)}$ dominates the
potential.  In this case, the transition temperature $T_{\rm tr}$ is given by
\begin{equation}
T_{\rm tr} \simeq m_{\phi(3/2)}. 
\end{equation}
Here we have approximated $|K|a_{\rm diff}\simeq 1$. 

In the case where there is $Q$-ball dominated era, the Hubble parameter
decreases with the temperature as
\begin{equation}
H\simeq  \left\{
\begin{array}{ll}
\dfrac{A}{A_T^{1/2}}\dfrac{T^4}{T_R^2M_G} & {\rm for} \ \ T>T_R, \\
A^{1/2}\dfrac{T^2}{M_G} & {\rm for} \ \ T_{\rm dom}<T<T_R, \\
A^{1/2} \dfrac{T^{3/2}T_{\rm dom}^{1/2}}{M_G} &{\rm for} \ \  T_p<T<T_{\rm dom}, \\
\dfrac{A}{A_{\rm dec}^{1/2}}\dfrac{T^4}{T_{\rm dec}^2M_G} & {\rm for} \ \ T_{\rm dec}<T<T_p, \\
A^{1/2}\dfrac{T^2}{M_G} & {\rm for} \ \ T<T_{\rm dec}.
\end{array}\right.
\end{equation}
Here $T_p\equiv(T_{\rm dec}^4 T_{\rm dom})^{1/5}$.  Then ${\rm min} \{
|\Gamma_{\rm evap}|, |\Gamma_{\rm diff}|\}/(QH)$ becomes the highest at
$T\simeq m_{\phi(3/2)}$.
Thus, the $Q$-balls are evaporated out before their decays if the
following condition is satisfied: 
\begin{equation}
Q<\left\{
\begin{array}{ll}
\dfrac{4 \pi a_{\rm diff} A_T^{1/2}}{A_{\phi(3/2)}}\dfrac{T_R^2M_G}{m_{\phi(3/2)}^3}   &{\rm for}  \ \  T_R<m_{\phi(3/2)} , \\
\dfrac{4\pi a_{\rm diff}}{A_{\phi(3/2)}^{1/2}}\dfrac{M_G}{m_{\phi(3/2)}}  &{\rm for}  \ \  T_{\rm dom}<m_{\phi(3/2)}< T_R,  \\
\dfrac{4 \pi a_{\rm diff}}{A_{\phi(3/2)}^{1/2}}\dfrac{M_G}{m_{\phi(3/2)}^{1/2}T_{\rm dom}^{1/2}}  &{\rm for} \ \  m_{\phi(3/2)}<T_{\rm dom}. 
\end{array}\right.\label{evgrq1}
\end{equation}
Here the subscript ``$\phi(3/2)$'' indicates that $A$ is estimated at $T=m_{\phi(3/2)}$.  

\subsection{Gauge-mediated case}

Here we consider the case where $V_{\rm gauge}$ dominates the potential.
In this case, the transition temperature $T_{\rm tr}$ is given by
\begin{equation}
T_{\rm tr} \simeq \left\{
\begin{array}{ll}
\sqrt{2}a_{\rm diff} M_F Q^{-1/4}   \ \ &{\rm for} \ \  T_{\rm tr}>m_\phi,  \\
(\sqrt{2}a_{\rm diff} m_\phi^2 M_F Q^{-1/4})^{1/3}   \ \ & {\rm for}  \ \ T_{\rm tr} < m_\phi. 
\end{array}\right. 
\end{equation}
These two temperatures coincide for $Q\simeq Q_{\rm cr} \equiv a^4 (M_F/m_\phi)^4$. 

Using the similar argument above, we have the condition that the $Q$-balls
are evaporated out before their decays,
\begin{equation}
Q< \left\{
\begin{array}{ll}
\left(\dfrac{2\sqrt{2}\pi}{A_T^{1/2}}\right)^{4/3}\left(\dfrac{T_R^2M_G}{m_\phi^2 M_F}\right)^{4/3} \ \ &{\rm for} \ \ T_R<m_\phi, T_{\rm tr}, \\
\left(\dfrac{2\sqrt{2}\pi}{A_T^{1/2}}\right)^{4/3}\left(\dfrac{M_G}{M_F}\right)^{4/3} \ \ &{\rm for} \ \ T_p, T_{\rm dom},  m_\phi <T_R, T_{\rm tr},   \\
\left(\dfrac{2\sqrt{2}\pi a_{\rm diff}^{2/3}}{A_T^{1/2}}\right)^{12/11}\left(\dfrac{M_G}{m_\phi^{2/3}M_F^{1/3}}\right)^{12/11} \ \ &  {\rm for} \ \  T_{\rm dec}<T_{\rm tr}< m_\phi, T_R,  \\
\dfrac{2\sqrt{2}\pi}{A_{\phi(3/2)}^{1/2}}\dfrac{T_R^2M_G}{m_\phi^2 M_F} \ \ & {\rm for} \ \  T_p<m_\phi<T_{\rm dom}<T_R<T_{\rm tr},  \\
\left(\dfrac{2\sqrt{2}\pi}{A_{\rm dec}^{1/2}}\right)^{4/3}\left(\dfrac{T_{\rm dec}^2M_G}{m_\phi^2 M_F}\right)^{4/3} \ \ & {\rm for} \ \ T_{\rm dec}<T_{\rm tr}<T_p.   
\end{array}\right.\label{evgaq1}
\end{equation}

In conclusion, $Q$-balls cannot be evaporated out before their decays
unless the charge $Q$ inside a $Q$-ball is small enough.

\section{The case with the zero-temperature potential}

\label{App:zero}

In this appendix, we show that the GWs from the $Q$-ball formation with
zero-temperature potential are too small to be detected.

\subsection{Gravity-mediated type}
\label{App:gravmed}

\begin{figure}[htbp]
 \begin{center}
  \includegraphics[width=9cm]{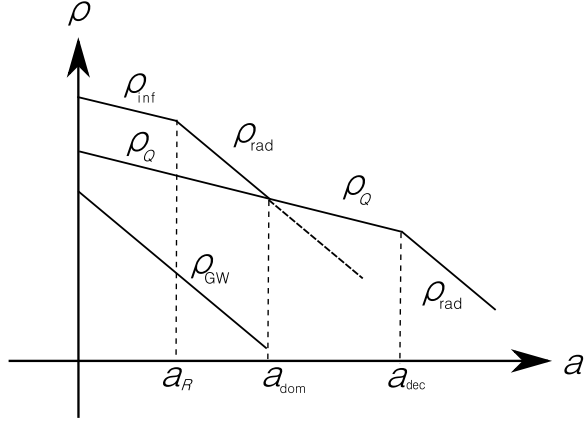}
 \end{center}
 \caption{ 
The time evolution of the energy density of each component is shown schematically. 
While $\rho_{\rm rad}$ and $\rho_{\rm GW}$ decrease in proportion to $a^{-4}$, $\rho_{\rm inf}$ in the inflaton oscillation era, and $\rho_Q$ decrease in proportion to $a^{-3}$
 \label{dilutionpic1}}
\end{figure}

First we consider the gravity-mediated or the ``new'' type $Q$-ball.  This
type is realized for the gravity or anomaly mediated SUSY
breaking model with $K<0$ or for the gauge-mediated SUSY-breaking model with
$V_{\rm grav(2)} (K<0)$.  In this type, the potential is dominated by
\begin{equation}
V_{\rm grav(2)} \simeq m_{\phi(3/2)}^2 \left[1+K\log\left(\dfrac{|\Phi|^2}{M_G^2} \right)\right]|\Phi|^2,  
\end{equation}
and from Eqs. \eqref{growmode} and \eqref{growrate}, the parameters
associated with the $Q$-ball formation are estimates as
\begin{equation}
\beta_{\rm gr}\simeq\frac{3}{4}m_{\phi(3/2)}|K|,  \ \ \frac{k_{\rm max}^2}{a^2} \simeq \frac{3}{2} m_{\phi (3/2)}^2|K|, \ \ {\rm and} \ \ H_*\simeq \frac{m_{\phi(3/2)}|K|}{2 \alpha}. \label{paragrz}
\end{equation}
Here, the factor $\alpha \simeq 30$ represents the dilution due to the
cosmic expansion.  Then, the amount of the GWs, $\Omega_{\rm GW}(k_{\rm
max}/(\pi a))$, at the $Q$-ball formation is estimated from
Eq.~\eqref{omggw*} as 
\begin{equation}
\Omega_{\rm GW}^*(\sqrt{6}m_\phi|K|^{1/2}/(2\pi))\simeq \frac{3}{16}\alpha^2 {\tilde \beta}^2\left(\frac{|K|}{\pi}\right)^3  \frac{\phi_{\rm osc}^4}{ M_G^4}. 
\end{equation} 
The baryon or lepton charge stored in a produced $Q$-ball is estimated as
\cite{kk01}
\begin{equation}
Q \simeq {\tilde \beta} \left(\frac{\phi_{\rm osc}}{m_{\phi(3/2)}}\right)^2, \label{grzq}
\end{equation}
where the numerical factor ${\tilde \beta} \simeq 6\times 10^{-3}$
represents the dilution due to the cosmic expansion.  Other properties
of the $Q$-balls are given by
\begin{align}
R^2&\simeq \frac{2}{m_{\phi(3/2)}^2|K|}, & \omega&\simeq m_{\phi(3/2)}, \notag \\
\phi_Q&\simeq \left(\frac{|K|}{\pi}\right)^{3/4} {\tilde \beta}^{1/2}\phi_{\rm osc}, & E_Q & \simeq \frac{1}{4}m_{\phi(3/2)}Q. 
\end{align}
Then, the average density of the $Q$-balls can be estimated as 
\begin{equation}
\rho_Q^* \simeq \frac{ m_{\phi(3/2)}^2 \phi_{\rm osc}^2{\tilde \beta}}{\eta}.  \\
\end{equation}
Here $\eta$ is a numerical factor of order unity.
Now we investigate the present properties of the GWs from the $Q$-ball
formation.  One should notice that there is no $Q$-ball transformation
different from the case when the thermal logarithmic potential dominates
the potential. Therefore, we have only to consider two cases depending on 
whether there is a $Q$-ball dominated era or not.  Figure
\ref{dilutionpic1} shows the schematic time evolution of the energy
density of each component.
The Hubble parameters at the $Q$-ball domination (if any) and the
$Q$-ball decay are given, respectively, by
\begin{align}
H_{\rm dom} &= \frac{16 \alpha^4 A_R^{1/2}{\tilde \beta}^4\phi_{\rm osc}^4T_R^2}{9\eta^4|K|^2M_G^5}, \label{hdomgrz}\\
H_{\rm dec} &=\frac{{\tilde \beta}^{-1}m_{\phi(3/2)}^3}{24 \pi \phi_{\rm osc}^2}. \label{hdecgrz}
\end{align}
Here we have used Eq.~\eqref{decgen} and the relations 
\begin{align}
\Omega_Q^* & \simeq \frac{4 \alpha^2 {\tilde \beta}^2 \phi_{\rm osc}^2}{3 \eta^2 |K|^2 M_G^2},  \\
H_{\rm dom} & \simeq \Omega_Q^{*2}H_R. \label{hdomr}
\end{align}  
Then, we have the condition for the $Q$-ball domination, 
\begin{equation}
\phi_{\rm osc} > \left(\frac{3}{128 \pi}\right)^{1/6} \frac{\eta^{2/3}|K|^{2/3}}{\alpha^{2/3} A_R^{1/12} {\tilde \beta}^{5/6}} \frac{M_G^{5/6}m_\phi^{1/2}}{T_R^{1/3}} \equiv \phi_{\rm osc}^c. \label{upphigrz}
\end{equation}
The present density parameter of the GWs from the $Q$-ball formation in
each case is given by
\begin{align}
\Omega_{\rm GW}^0&\simeq\left\{
\begin{array}{ll}
\dfrac{1}{192} \left(\dfrac{3}{\pi}\right)^{2/3} \dfrac{|K|^5 {\tilde \beta}^{-4/3} \eta^{8/3}}{\pi^3} \dfrac{m_\phi^{4/3}}{M_G^{4/3}} \dfrac{a_{\rm eq}}{a_0} & \text{(with the }Q\text{-ball domination)}, \\ 
\dfrac{ 2^{2/3}}{12} \dfrac{\alpha^{8/3}{\tilde \beta}^2|K|^{7/3} A_R^{1/3}}{\pi^3}\dfrac{\phi_{\rm osc}^4 T_R^{4/3}}{M_G^{14/3}m_\phi^{2/3}}\dfrac{a_{\rm eq}}{a_0} & \text{(without the }Q\text{-ball domination)}. 
\end{array}\right.
\end{align}
Here we have used Eqs.~\eqref{dilute}, \eqref{paragrz}, \eqref{hdomgrz} and \eqref{hdecgrz}. 
Then, from Eq.~\eqref{upphigrz}, we have the upper bound on the present density parameter of GWs, 
\begin{equation}
\Omega_{\rm GW}^0 \leq \dfrac{1}{192} \left(\dfrac{3}{\pi}\right)^{2/3} \dfrac{|K|^5 {\tilde \beta}^{-4/3} \eta^{8/3}}{\pi^3} \dfrac{m_\phi^{4/3}}{M_G^{4/3}} \dfrac{a_{\rm eq}}{a_0} \simeq 10^{-25}, 
\end{equation}
which is too small to be detected by the future detectors. 

Here we comment on the possibility of the evaporation of $Q$-balls before
their decay.  From Eq. \eqref{evgrq1}, we find that the $Q$-balls are
evaporated out before their decays if the following condition is
satisfied: 
\begin{equation}
Q<\frac{4\pi a}{A_{\phi(3/2)}^{1/2}}\frac{M_G}{m_{\phi(3/2)}}, 
\end{equation}
which yields, using Eq.~\eqref{grzq}, 
\begin{equation}
\phi_{\rm osc}<\left(\frac{4 \pi a }{A_{\phi(3/2)}^{1/2}{\tilde \beta}}\right)^{1/2} m_{\phi(3/2)}^{1/2}M_G^{1/2} \lesssim 10^{12} {\rm GeV}. 
\end{equation}
Such a small initial field value does not produce significant amounts of
GWs and have the present amplitude of GWs is rather small.  Thus, our
conclusion is unchanged even if we take into account of the possibility 
that the $Q$-balls are evaporated out before their decays.

\subsection{Gauge-mediated type}

\label{subsec:gauge}

Next we consider the gauge-mediated type $Q$-balls.  In this type, the
potential is dominated by
\begin{equation}
V_{\rm gauge} \simeq M_F^4 \left(\log \dfrac{|\Phi|^2}{M_S^2} \right)^2. 
\end{equation}
This potential is quite similar to the thermal logarithmic potential. 
Then, the parameters associated with the $Q$-ball formation are given by
\begin{equation}
\frac{k_{\rm max}^2}{a^2}\simeq \frac{3}{2}\frac{M_F^4}{\phi_{\rm osc}^2},  \ \ \beta_{\rm gr} \simeq \frac{M_F^2}{\sqrt{2}\phi_{\rm osc}} \ \ {\rm and} \ \ 
H_*=\frac{1}{\alpha} \frac{M_F^2}{\phi_{\rm osc}},  \label{gmgwgauge}
\end{equation}
which yield the amount of the GWs $\Omega_{\rm GW}(k_{\rm max}/(\pi a))$
at the $Q$-ball formation from Eq.~\eqref{omggw*},
\begin{equation}
\Omega_{\rm GW}^*(\sqrt{6}M_F^2/(2 \pi \phi_{\rm osc}))\simeq
 \frac{\alpha^2}{54}({\bar \epsilon}\beta)\frac{\phi_{\rm osc}^4}{
 M_G^4}. 
\end{equation}
Other properties of the produced $Q$-balls are obtained by replacing
$\alpha_g^{1/2} T_{\rm osc}$ by $M_F$ in Eq. (\ref{qthermal}).

Now we evaluate the present properties of the GWs from the $Q$-ball
formation.  Note that there is no $Q$-ball transformation in this type and
hence the cosmic history depends on whether their is $Q$-ball dominated
era or not, as is the case with the gravity-mediated type (Appendix
\ref{App:gravmed}). The Hubble parameters at the $Q$-ball domination
(if any) and the $Q$-ball decay are, respectively, given by
\begin{align}
H_{\rm dom} &= \frac{\alpha^4 \phi_{\rm osc}^4}{9 \eta^2 M_G^4} A_R^{1/2}({\bar \epsilon}\beta)^{3/2}\frac{T_R^2}{M_G}, \\
H_{\rm dec} &= \frac{\sqrt{2}\pi^2}{48}({\bar \epsilon}\beta)^{-5/4}\frac{M_F^6}{\phi_{\rm osc}^5}, 
\end{align}
which give the condition for the $Q$-ball domination, 
\begin{equation}
\phi_{\rm osc}>\left(\frac{3\sqrt{2}\pi^2}{16}\right)^{1/4}({\bar \epsilon}\beta)^{-11/36} \frac{\eta^{2/9}}{\alpha^{4/9}A_R^{1/18}}\frac{M_F^{2/3}M_G^{5/9}}{T_R^{2/9}} \equiv \phi_{\rm osc}^c. 
\end{equation}
Here we have used Eqs.~\eqref{decgen} and \eqref{hdomr}.  Using
Eq.~\eqref{dilute}, the present density parameter of GWs from $Q$-ball
formation is estimated as
\begin{align}
\Omega_{\rm GW}^0&\simeq\left\{
\begin{array}{ll}
\dfrac{1}{216}\left(\dfrac{3\pi^2}{\sqrt{2}}\right)^{2/3}({\bar \epsilon}\beta)^{-5/6} \alpha^4 \eta^{4/3}\dfrac{M_F^{8/3}}{M_G^{4/3}\phi_{\rm osc}^{4/3}} \dfrac{a_{\rm eq}}{a_0} & \text{ (with  the  }Q\text{-ball domination)}, \\ 
\dfrac{\alpha^{8/3}}{54}({\bar \epsilon}\beta) A_R^{1/3}\dfrac{\phi_{\rm osc}^{14/3}T_R^{4/3}}{M_G^{14/3}M_F^{4/3}}\dfrac{a_{\rm eq}}{a_0} & \text{ (without  the }\text{Q-ball domination)}.
\end{array}\right.
\end{align}
The condition that $\phi_Q= ({\bar \epsilon}\beta)^{1/4} \phi_{\rm osc}>M_S=M_F^2 /m_\phi$ yields the upper bound on $\Omega_{\rm GW}^0$, 
\begin{equation}
\Omega_{\rm GW}^{0,{\rm max}}\simeq \frac{1}{216}\left(\frac{3\pi^2}{\sqrt{2}}\right)^{2/3}({\bar \epsilon}\beta)^{-5/6} \alpha^4 \eta^{4/3}\frac{m_\phi^{4/3}}{M_G^{4/3}}\dfrac{a_{\rm eq}}{a_0} \simeq 10^{-21}. 
\end{equation}
Again, the GWs generated by the formation of the
gauge-mediated type $Q$-balls are too small to be detected by the
next-generation GW detectors.

Finally, we give the condition that the $Q$-balls are evaporated out
before their decays,
\begin{equation}
Q\lesssim \left(\frac{M_G}{M_F}\right)^{4/3} \Leftrightarrow \phi_{\rm osc}\lesssim M_F^{4/3}M_G^{-1/3}, 
\end{equation}
which yields $\phi_{\rm osc} \lesssim 10^{8}$ GeV since $M_F<10^{10}$
GeV.  Such a small initial field value does not produce significant
amounts of GWs and hence the present amplitude of GWs is rather small.
Thus, our conclusions are unchanged even if we take the $Q$-ball
evaporation into account.

\subsection{Delayed type}

Next, we consider the delayed type $Q$-balls.  In the gauge-mediated 
SUSY-breaking model, the effective potential for the AD fields is given by
the summation of $V_{\rm grav2}$, $V_{\rm gauge}$, and $V_{\rm thermal}$.
If the AD field starts its oscillation where the potential is dominated
by $V_{\rm grav2}$ with $K>0$, $Q$-balls are not formed and the AD field
falls down along the potential.  Then, $V_{\rm thermal}$ or $V_{\rm
gauge}$ dominates the potential of the AD field at a critical
temperature or a critical field value, which induces the formation of
$Q$-balls, called the delayed type $Q$-balls.
Thus, there are two cases depending which potential term is responsible
for the $Q$-ball formation.

\subsubsection{$V_{\rm gauge}$ driven $Q$-ball formation}

First we consider the case where $V_{\rm gauge}$ drives the $Q$-ball
formation.  In this case, when $\phi=\phi_{eq}\equiv M_F^2/m_{3/2}$,
$Q$-balls are formed.  The Hubble parameter at the $Q$-ball formation is
given by
\begin{equation}
H_*\simeq \frac{M_F^2}{\phi_{\rm osc}}, 
\end{equation}
where $\phi_{\rm osc}$ is the field value at the onset of the
oscillation of the AD field and we have used the relation $\phi \propto
H$ during the oscillation of the AD field.
The condition that $V_{\rm gauge}$ dominates the potential before
$V_{\rm thermal}$ is given by $T_R<M_F$.  From Eqs.~\eqref{growmode} and \eqref{growrate}, the parameters associated with
the $Q$-ball formation are given by
\begin{equation}
\beta_{\rm gr}\simeq \frac{m_{3/2}}{\sqrt{2}} \ \ {\rm and} \ \ \frac{k^2_{\rm max}}{a^2}\simeq m_{3/2}^2, 
\end{equation}
which yield the amount of the GWs, $\Omega_{\rm GW}^*(k_{\rm max}/(\pi
a))$, at the $Q$-ball formation from Eq.~\eqref{omggw*},
\begin{equation}
\Omega_{\rm GW}^*(m_{3/2}/\pi)\simeq \frac{\phi_{\rm osc}^2M_F^4}{54M_G^4m_{3/2}^2}. 
\end{equation}
The baryon or lepton charges stored in a $Q$-ball are given by \cite{kk01} 
\begin{equation}
Q \simeq \left(\frac{\phi_{eq}}{M_F}\right)^4 \simeq \left(\frac{M_F}{m_{3/2}}\right)^4. 
\end{equation}
Here the numerical factor $\beta$ [defined in \eqref{chthermal}] is almost unity since the $Q$-balls are
formed very quickly so that the cosmic expansion is negligible.  Other
properties of $Q$-balls are estimated as,
\begin{align}
R&\simeq \frac{Q^{1/4}}{\sqrt{2}M_F} \simeq (\sqrt{2}m_{3/2})^{-1}, & \omega&\simeq \frac{\sqrt{2}\pi M_F}{Q^{1/4}}\simeq \sqrt{2} \pi m_{3/2}, \notag \\
\phi_Q&\simeq  M_F Q^{1/4} \simeq \frac{M_F^2}{m_{3/2}},  &E_Q&\simeq \frac{4\pi\sqrt{2}}{3}M_FQ^{3/4} .  \label{qdelay}
\end{align}
The energy density of $Q$-balls at the $Q$-ball formation is given by
\begin{equation}
\rho_Q^*\simeq E_Q (k_{\rm max}^3/a^3) \simeq M_F^4 . 
\end{equation} 

Then, we evaluate the present properties of the GWs from the $Q$-ball
formation.  Again, there is no $Q$-ball transformation and hence
the cosmic history depends on whether there is the $Q$-ball dominated era or
not.  The Hubble parameters at the $Q$-ball domination (if any) and the
$Q$-ball decay are, respectively, given by
\begin{align}
H_{\rm dom} &= \frac{A_R^{1/2}\phi_{\rm osc}^4 T_R^2}{9M_G^5}, \\
H_{\rm dec} &= \frac{\pi^2}{24 \sqrt{2}}\frac{m_{3/2}^5}{M_F^4}. 
\end{align}
Here we have used the relations, 
\begin{align}
\Omega_Q^* & \simeq\frac{\phi_{\rm osc}^2}{3M_G^2},  \\
H_{\rm dom} & \simeq \Omega_Q^{*2}H_R. 
\end{align}  
Thus, the condition for the $Q$-ball domination is given by 
\begin{equation}
\phi_{\rm osc}>A_R^{-1/8}\left(\frac{3 \pi^2}{8\sqrt{2}}\right)^{1/4}\frac{m_{3/2}^{5/4}M_G^{5/4}}{T_R^{1/2}M_F}. 
\end{equation}
{}From Eq.~\eqref{dilute}, the present density parameter of GWs from $Q$-ball formation is given by
\begin{align}
\Omega_{\rm GW}^0&\simeq\left\{
\begin{array}{ll}
\dfrac{1}{216} \left(\dfrac{3\pi^2}{\sqrt{2}}\right)^{2/3} \dfrac{m_{3/2}^{4/3}}{M_G^{4/3}} \dfrac{a_{\rm eq}}{a_0} & \text{(with  the }Q\text{-ball  domination)}, \\ 
\dfrac{A_R^{1/3}}{54} \dfrac{\phi_{\rm osc}^{8/3} T_R^{4/3}M_F^{8/3}}{M_G^{14/3}m_{3/2}^2}\dfrac{a_{\rm eq}}{a_0} & \text{(without the }Q\text{-ball  domination)}. 
\end{array}\right.
\end{align}
Considering the condition, that $m_{3/2}<10 {\rm GeV}$ for the
gauge-mediated SUSY-breaking model, we have the constraint on
$\Omega_{\rm GW}^0$,
\begin{equation}
\Omega_{\rm GW}^0 <  10^{-24}, 
\end{equation}
which is too small for the detection.  This conclusion applies also for the
second $Q$-ball formation discussed in Sec. \ref{subsubsec:C}.

Here we comment on the $Q$-ball evaporation. 
{}From Eqs. \eqref{evgaq1}, the $Q$-balls are evaporated out before their decays if the following condition is satisfied: 
\begin{equation}
Q\lesssim \left(\frac{M_G}{M_F}\right)^{4/3} \Leftrightarrow M_F\lesssim M_G^{1/4}m_{3/2}^{3/4}.   
\end{equation}
Thus, we have the inequality with respect to the amount of the GWs at the
$Q$-ball formation,
\begin{equation}
\Omega_{\rm GW}^*<\frac{\phi_{\rm osc}^2m_{3/2}}{54 M_G^3}<10^{-16}. 
\end{equation}
Note that this amount of GWs is further diluted at least after
matter-radiation equality.  Therefore we conclude that even if we take
into account the possibility that the $Q$-balls are evaporated out before
their decays, the amount of the present GWs is too small to be detected
by the next-generation detectors.

\subsubsection{$V_{\rm thermal}$ driven $Q$-ball formation}

Next we consider the case  where $V_{\rm thermal}$ drives the $Q$-ball formation. 
In this case, when $T=T_*$ and $\phi_*\simeq \alpha_g T_*^2/m_{3/2}$, $Q$-balls are formed. 
Note that the reheating takes place after the $Q$-ball formation because we have the following inequality: 
\begin{equation}
T_* \simeq \left(\frac{\alpha_g^2}{A_T^2}\right)^{1/4}\left(\frac{M_G}{\phi_{\rm osc}}\right)^{1/2}T_R >T_R. 
\end{equation} 
In this case, we have two constraints on the reheating temperature. 
One is the condition that $V_{\rm grav2}$ dominates $V_{\rm thermal}$ at the onset of the AD field oscillation, 
\begin{equation}
T_R<\frac{A_T^{1/4}}{\alpha_g}\frac{m_{3/2}^{1/2}}{M_G^{1/2}}\phi_{\rm osc} \equiv T_R^{c,{\rm gr}}. \label{condtrii}
\end{equation}
Another condition is that $V_{\rm thermal}$ dominates $V_{\rm gauge}$ at the $Q$-ball formation, 
\begin{equation}
T_R>\frac{A_T^{1/4}}{\alpha_g}\frac{\phi_{\rm osc}^{1/2}}{M_G^{1/2}}M_F \equiv T_R^{c,{\rm th}}. 
\end{equation}

The properties associated with the $Q$-ball formation are given by 
\begin{equation}
\beta_{\rm gr}\simeq \frac{m_{3/2}}{\sqrt{2}} \ \ {\rm and} \ \ \frac{k^2_{\rm max}}{a^2}\simeq m_{3/2}^2, 
\end{equation}
which gives the amount of the GWs, $\Omega_{\rm GW}^*(k_{\rm max}/(\pi a))$,  at the $Q$-ball formation, 
\begin{equation}
\Omega_{\rm GW}^*(m_{3/2}/\pi)\simeq \frac{\alpha_g^4  T_R^4}{54A_T M_G^2m_{3/2}^2}. 
\end{equation}
Here we have used the following relations:  
\begin{equation}
H_* \simeq \alpha_g \frac{T_*^2}{\phi_{\rm osc}}\simeq \frac{\alpha_g^2}{A_T^{1/2}}\left(\frac{T_R}{\phi_{\rm osc}}\right)^2 M_G. 
\end{equation}
The baryon or lepton charges stored in a $Q$-ball are given by \cite{kk01}, 
\begin{equation}
Q \simeq \left(\frac{\phi_*}{\alpha_g^{1/2}T_*}\right)^4 \simeq \frac{\alpha_g^4}{A_T}\left(\frac{M_G}{\phi_{\rm osc}}\right)^2 \left(\frac{T_R}{m_{3/2}}\right)^4. 
\end{equation}
Here the numerical factor $\beta$ is almost unity since  the $Q$-balls are formed very quickly so that the cosmic expansion is negligible at the formation in this case. 
Other properties of $Q$-balls are evaluated as 
\begin{align}
R&\simeq \frac{Q^{1/4}}{\sqrt{2}\alpha_g^{1/2}T}, & \omega&\simeq \frac{\sqrt{2}\pi \alpha_g^{1/2}T}{Q^{1/4}}, \notag \\
\phi_Q&\simeq  \alpha_g^{1/2}T Q^{1/4} ,  &E_Q&\simeq \frac{4\pi\sqrt{2}}{3}\alpha_g^{1/2}TQ^{3/4} .  \label{qdelayth}
\end{align}
At the $Q$-ball formation, the energy density of $Q$-balls is given by 
\begin{equation}
\rho_Q^*\simeq \alpha_g^2 T_*^4 \simeq  \frac{\alpha_g^4 }{A_T}\left(\frac{M_G}{\phi_{\rm osc}}\right)^2 T_R^4, 
\end{equation}
and it decreases with the temperature as
\begin{equation}
\rho_Q \propto A^2T^9. 
\end{equation}

Next, we follow the cosmic history after the $Q$-ball formation and
evaluate the present properties of the GWs from the $Q$-ball formation in
this case.  The properties of $Q$-balls are changed into those of the
gauge-mediated type $Q$-balls, when $V_{\rm gauge}$ dominates $V_{\rm
thermal}$ at $T\simeq T_C \equiv \alpha_g^{-1/2}M_F$.  In the same way
as the cases discussed in Sec. \ref{subsec:detectgw}, and
Sec. \ref{gaugecase}, where the thermal logarithmic potential dominates
the effective potential, we have four possibilities of the cosmic
history after the $Q$-ball formation.  They are characterized by the
critical temperatures, $T_R^{c1}$ and $T_R^{c2}$. In this case, they are
summarized in Table \ref{Table:5}.
\begin{table}[htbp]
\begin{center}
\begin{tabular}{l|l}
\hline
Case 1 & \ $T_R>{\rm max} \{T_R^{c1}(\phi_{\rm osc}), T_R^{c2}(\phi_{\rm osc})\}$  \\ \hline
Case 2 & \ $T_R^{c1}(\phi_{\rm osc})<T_R<T_R^{c2}(\phi_{\rm osc})$ \\
\hline
Case 3 & \ $T_R^{c2}(\phi_{\rm osc})<T_R<T_R^{c1}(\phi_{\rm osc})$ \\ \hline
Case 4 & \ $T_R<{\rm min} \{T_R^{c1}(\phi_{\rm osc}), T_R^{c2}(\phi_{\rm osc})\}$ \\ \hline
\end{tabular} 
\end{center}
\caption{The conditions of four cases of the cosmic history \label{Table:5}}
\end{table}

The critical temperatures are given by
\begin{align}
T_R^{c1} & = \alpha_g^{-1/2} M_F, \\
T_R^{c2} & = \left(\frac{3\pi^2}{8\sqrt{2}}\right)^{1/5}  \dfrac{\alpha_g^{3/5} A_T^{3/20}}{A_{\rm dom}^{1/10}}\dfrac {m_{3/2} M_G^{7/10}} {\phi_{\rm osc}^{1/2} M_F^{1/5}}.  
\end{align}
Here we have used the Hubble parameters at the $Q$-ball domination
(if any) and the $Q$-ball decay given by
\begin{align}
H_{\rm dom}&\simeq \dfrac{A_{\rm dom}^{1/2}A_T^{1/2}}{9 \alpha_g^2}\dfrac{\phi_{\rm osc}^5M_F^2}{M_G^6}, \\
H_{\rm dec}&\simeq \dfrac{\pi^2}{24\sqrt{2}}\dfrac{A_T^{5/4}}{\alpha_g^5}\dfrac{\phi_{\rm osc}^{5/2}m_{3/2}^5 M_F}{M_G^{5/2}T_R^5},  
\end{align}
Then, from Eq.~\eqref{dilute} the present density parameter of GWs from $Q$-ball formation is estimated as 
\begin{align}
\Omega_{\rm GW}^0&\simeq\left\{
\begin{array}{ll}
\dfrac{1}{216} \left(\dfrac{3 \pi^2}{\sqrt{2}}\right)^{2/3} \dfrac{\alpha_g^{2/3}A_T^{1/6}}{ A_{\rm dom}^{1/3}} \dfrac{T_R^{2/3} m_{3/2}^{4/3}}{M_G M_F^{2/3} \phi_{\rm osc}^{1/3}} \dfrac{a_{eq}}{a_0} & \quad {\rm (Case \ 1,3)}, \\ 
\dfrac{\alpha_g^{8/3}A_T{4/3}}{54} \dfrac{T_R^4 \phi_{\rm osc}^{4/3}}{M_G^{10/3}m_{3/2}^2 }\dfrac{a_{eq}}{a_0} & \quad {\rm (Case \ 2,4)}. \\
\end{array} \right. 
\end{align}
The condition Eq. \eqref{condtrii} yields the upper limit of the present
amount of the GWs in Case 1 and Case 3,
\begin{equation}
\Omega_{\rm GW}^{0, {\rm max}} \simeq 
\dfrac{1}{216}\left(\dfrac{3 \pi^2}{\sqrt{2}}\right)^{2/3} \dfrac{A_T^{1/3}}{ A_{\rm dom}^{2/3}} \frac{m_{3/2}^{5/3}\phi_{\rm osc}^{1/3}}{M_G^{4/3}M_F^{2/3}}\frac{a_{eq}}{a_0} \leq 10^{-24}, 
\end{equation}
which is too small to be detected.  Here we have used the fact that
$\alpha_g \simeq 0.1, \phi_{\rm osc}<M_G, m_{3/2}<10 {\rm GeV},
M_F>10^4{\rm GeV}$, and $a_{eq}/a_0\simeq 3\times 10^{-4}$. $g_*$ and
$g_{*s}$ are evaluated in the context of MSSM.  Moreover, in Case
2 and Case 4, the constraint on the reheating temperature $T_R<T_R^{c2}$
strongly constrains the upper bound of $\Omega_{GW}^0$ and hence the
present amount of the GWs cannot be larger than that in Case 1 and Case
3. Therefore we conclude that the GWs generated by the formation of this
type of $Q$-balls are too small to be detected by the next-generation GW
detectors.

Here we comment on the effect of the $Q$-ball evaporation.  From
Eq. \eqref{evgaq1}, the $Q$-balls are evaporated out before their decays
if the following condition is satisfied:
\begin{equation}
Q\lesssim \left(\frac{M_G}{M_F}\right)^{4/3} \Leftrightarrow T_R<\frac{A_T^{1/4}}{\alpha_g}\frac{\phi_{\rm osc}^{1/2}m_{3/2}}{M_G^{1/6}M_F^{1/3}},  
\end{equation}
which gives the inequality for the amount of the GWs at the $Q$-ball formation, 
\begin{equation}
\Omega_{\rm GW}^*\lesssim\frac{1}{54}\frac{m_{3/2}^2 \phi_{\rm osc}^2}{M_G^{8/3}M_F^{4/3}}. 
\end{equation}
Considering the fact that $m_{3/2}\lesssim 10$ GeV, $\phi_{\rm osc}\lesssim M_G$, and $M_F \gtrsim 10^4$ GeV, we have the upper limit 
on $\Omega_{\rm GW}^*$, 
\begin{equation}
\Omega_{\rm GW}^*<10^{-16}, 
\end{equation}
which is further diluted at least after matter-radiation equality.
Therefore we conclude that even if we take into account the possibility
that the $Q$-balls are evaporated out before their decays, the amount of
the present GWs is too small to be detected by the next-generation
detectors.

\subsection{Negative thermal log type}

Finally, we consider the case with negative thermal logarithmic
potential.  So far, we assumed that the contribution of the thermal
logarithmic potential is positive.  However, it is possible for the term
to be negative and hence there can be another type of $Q$-balls
\cite{Kasuya:2003yr}.

We then consider this type of $Q$-balls. If the temperature after
inflation is sufficiently high, the AD field is trapped in the potential
minimum, $\phi\simeq(\alpha_g T^2 M^{n-3})^{1/(n-1)}$, determined by the
balance between the nonrenormalizable $F$-term and the negative thermal
log term, rather than the negative Hubble mass term. In the gravity or
anomaly mediated SUSY-breaking model, this potential minimum exists
until the thermal correction to the potential turns to the thermal mass
term.  At that time, the AD field starts oscillating around the origin
by the thermal mass term so that it decays quickly. Thus, $Q$-balls
are not formed in this case. On the other hand, in the gauge-mediated
SUSY-breaking model, this potential minimum vanishes when
$T=\alpha_g^{-1/2}M_F$. In this case, the AD field starts oscillating
{}from $\phi_{\rm osc}\simeq (M_F^2T^2 M^{n-3})^{1/(n-1)}$ by $V_{\rm
gauge}$ so that $Q$-balls are formed.  Hereafter, we consider such a type
of $Q$-balls.

In this case, the parameters associated with the $Q$-ball formation are given by, 
\begin{equation}
\frac{k_{\rm max}^2}{a^2}\simeq \frac{3}{2}\frac{M_F^4}{\phi_{\rm osc}^2},  \ \ \beta_{\rm gr} \simeq \frac{M_F^2}{\sqrt{2}\phi_{\rm osc}} \ \ {\rm and} \ \ 
H_*\simeq\left\{
\begin{array}{ll}
\dfrac{A_*^{1/2}}{\alpha_g} \dfrac{M_F^2}{M_G} & {\rm for} \ T_R>\alpha_g^{-1/2} M_F, \\ 
\dfrac{A_*}{\alpha_g^2 A_R^{1/2}} \dfrac{M_F^4}{M_G T_R^2}& {\rm for} \ T_R<\alpha_g^{-1/2} M_F. 
\end{array}\right.   \label{negthpara}
\end{equation}
The properties of the $Q$-balls are estimated as 
\begin{equation}
Q\simeq \left(\frac{\phi_{\rm osc}}{M_F}\right)^4, \ \ \phi_Q\simeq \phi_{\rm osc},  \ \ {\rm and} \ \  \rho_Q^* \simeq M_F^4  . 
\end{equation}
Then, the amounts of the GWs, $\Omega_{\rm GW}^*$, at the $Q$-ball formation are given by
\begin{equation}
\Omega_{\rm GW}^* \simeq \left\{
\begin{array}{ll}
\dfrac{\alpha_g^2}{54 A_*}\left(\dfrac{\phi_{\rm osc}}{M_G}\right)^2 & {\rm for} \ T_R>\alpha_g^{-1/2} M_F, \\
\dfrac{\alpha_g^2 A_R}{54 A_*^2}\left(\dfrac{\phi_{\rm osc}}{M_G}\right)^2 \left(\dfrac{T_R}{\alpha_g^{-1/2}M_F}\right)^4 & {\rm for} \ T_R<\alpha_g^{-1/2} M_F. 
\end{array}\right.
\end{equation}
Note that since the $Q$-balls are formed quickly, there are no dilution factors. 

Next, we follow the cosmic history after the $Q$-ball formation and
evaluate the present properties of the GWs from the $Q$-ball formation.
In this case, there are four possibilities of the cosmic history.  They
are classified by two criterions.  One is whether the $Q$-ball dominated
era exists or not and the other depends which first takes place, the
reheating after the inflaton decay or the beginning of the oscillation
of the AD field.

The Hubble parameters at the $Q$-ball domination 
and at the $Q$-ball decay, $H_{\rm dec}$ are, respectively, given by 
\begin{align}
H_{\rm dom}&\simeq \left\{
\begin{array}{ll}
\dfrac{\alpha_g^3}{9A_*^{3/2}} \dfrac{M_F^2}{M_G}  & {\rm for} \ T_R>\alpha_g^{-1/2} M_F, \\ 
\dfrac{\alpha_g^{8}A_R^{3/2}}{9A_*^{3}} \dfrac{ T_R^{10}}{M_G M_F^{8}}  &  {\rm for} \ T_R<\alpha_g^{-1/2} M_F,   
\end{array} \right. \\
H_{\rm dec}&=\frac{\sqrt{2} \pi^2}{48}\frac{M_F^6}{\phi_{\rm osc}^5},   
\end{align}
which gives the condition for the $Q$-ball domination, 
\begin{equation}
\phi_{\rm osc}> \left \{
\begin{array}{ll}
\left(\dfrac{3\sqrt{2}\pi^2}{16}\right)^{1/5} \dfrac{A_*^{3/10}}{\alpha_g^{3/5}}M_G^{1/5}M_F^{4/5} & {\rm for} \ T_R>\alpha_g^{-1/2} M_F, \\ 
\left(\dfrac{3\sqrt{2}\pi^2}{16}\right)^{1/5} \dfrac{A_R^{3/5}}{\alpha_g^{8/5}A_*^{3/10}} \dfrac{M_G^{1/5}M_F^{14/5}}{T_R^2}&  {\rm for} \ T_R<\alpha_g^{-1/2} M_F. 
\end{array}\right. 
\end{equation}
The present density parameter of the GWs from the $Q$-ball formation is given by 
\begin{align}
\Omega_{\rm GW}^0&\simeq\left\{
\begin{array}{ll}
\dfrac{1}{216} \left(\dfrac{3 \pi^2}{\sqrt{2}}\right)^{2/3} \dfrac{M_F^{8/3}}{M_G^{4/3}\phi_{\rm osc}^{4/3}} \dfrac{a_{eq}}{a_0} & \quad \text{ with  the   }Q\text{-ball   domination},  \\ 
\dfrac{\alpha_g^2}{54A_*}\left(\dfrac{\phi_{\rm osc}}{M_G}\right)^2 \dfrac{a_{eq}}{a_0} & \quad \text{without  the   Q-ball   domination and } T_R>\alpha_g^{-1/2} M_F,  \\ 
\dfrac{\alpha_g^2A_R}{54A_*^2}\left(\dfrac{\phi_{\rm osc}}{M_G}\right)^2\left(\dfrac{T_R}{M_F}\right)^{20/3} \dfrac{a_{eq}}{a_0} & \quad \text{ without  the }Q\text{-ball  domination and } T_R<\alpha_g^{-1/2} M_F. 
\end{array} \right. 
\end{align}
Then, the upper bound on $\Omega_{\rm GW}^0$ can be obtained  by the same
consideration as the case discussed in Appendix \ref{subsec:gauge}
and hence are given by $\Omega_{\rm GW}^{0,{\rm max}}\simeq 10^{-24}$.
Moreover, the possibility of the $Q$-ball evaporation does not change
the conclusion as is the case in Appendix \ref{subsec:gauge}.
Therefore we conclude that the GWs generated by the formation of this
type of $Q$-balls are too small to be detected by the next-generation GW
detectors.

We conclude that the GWs from the $Q$-ball formation can hardly
be detected by the future detectors in the case where the oscillation of
the AD field is driven by the zero-temperature potential terms.


\end{document}